\documentclass[aip,amsmath,amssymb,reprint]{revtex4-1}

\usepackage{graphicx}
\usepackage{dcolumn}
\usepackage{bm}

\usepackage[T1]{fontenc}
\usepackage{mathptmx}
\usepackage{amsmath}
\usepackage{amssymb}
\usepackage{mathptmx}
\usepackage{floatrow}
\usepackage{caption}
\usepackage{subfig}
\usepackage{multirow}
\usepackage{makecell}
\usepackage{adjustbox}
\usepackage{xcolor}
\usepackage{textcomp}

\draft 

\begin{document}

\preprint{Paper2}

\title{Phase space reconstruction from a biological time series. \\A PhotoPlethysmoGraphic signal a case study} 

\author{J. de Pedro-Carracedo}
\email{javier.depedro@uah.es}
\affiliation{University of Alcal\'a (UAH), Computer Engineering Department, Alcal\'a de Henares (Madrid), 28871 Spain} 
\author{A.M. Ugena}
\email{anamaria.ugena@upm.es}
\affiliation{Technical University of Madrid (UPM), Departamento de Matem\'atica aplicada a las Tecnolog\'ias de la Informaci\'on, Madrid, 28040 Spain}
\author{A.P. Gonzalez-Marcos}
\email{anapilar.gonzalez@upm.es}
\affiliation{Technical University of Madrid (UPM), Photonic Technology and Bioengineering Department, Madrid, 28040 Spain}

\date{\today}

\begin{abstract}
In the analysis of biological time series, the state space comprises a framework for the study of systems with presumably deterministic properties. However, a physiological experiment typically captures an observable, or, in other words, a series of scalar measurements that characterize the temporal response of the physiological system under study; the dynamic variables that make up the state of the system at any time are not available. Therefore, only from the acquired observations should state vectors reconstructed in order to emulate the different states of the underlying system. It is what is known as the reconstruction of the state space, called phase space in real-world signals, for now only satisfactorily resolved using the method of delays. Each state vector consists of $m$ components, extracted from successive observations delayed a time $\tau$. The morphology of the geometric structure described by the state vectors, as well as their properties, depends on the chosen parameters $\tau$ and $m$. The real dynamics of the system under study is subject to the correct determination of the parameters $\tau$ and $m$. Only in this way can be deduced characteristics with true physical meaning, revealing aspects that reliably identify the dynamic complexity of the physiological system. The biological signal presented in this work, as a case study, is the \textbf{P}hoto\textbf{P}lethysmo\textbf{G}raphic (PPG) signal. We find that $m$ is five for all the subjects analyzed and that $\tau$ depends on the time interval in which it evaluates. The H\'enon map and the Lorenz flow are used to facilitate a more intuitive understanding of applied techniques. \end{abstract}


\maketitle 

\section{Introduction}\label{sec:IN}

Dynamic systems, as could be any physiological system, are mathematically characterized by differential equations. The identification of the simplest physiological model, to describe the physiological temporal evolution, requires specifying the minimum dimension of the dynamic system, that is, the number of dynamic variables or differential equations involved in the evolution of the system. The dynamic variables compose the components of a state vector, in the state space\footnote{The state space comprises a coordinate system with as many coordinates as dynamic variables the system presents. Therefore, space allows embracing the set of all the possible states of the system.}, that describes the dynamics of the system. At each time instant, the state vector is in a different position (an isolated point in the state space); the chronological evolution of these points draws up a trajectory in the state space. When the trajectory extends to infinity, it is known as an orbit\cite{Strogatz1994}. 

The time evolution of each dynamic variable involves the most natural way to characterize any dynamic system; this is nothing but the usual representation of the value of each dynamic variable as a function of time\cite{Baker1996}. Another way to describe a dynamical system graphically is to replace the time-independent variable, typically, the variable time $t$, with another dynamic variable of the system\cite{Ott2002,Devaney2018}. In this case, each point on the graph, in a two-dimensional coordinate system, represents the system state in a given time instant; with a third dynamic variable, a three-dimensional coordinate system is available, the highest possible visual capacity. In a hypothetical abstract space, the coordinate system must cover all dynamic variables of the system under study. Each point of the theoretical graph represents a system state at a given time instant. This graphic construct designates what is called the state space, where each point denotes a state vector. If the dynamic variables that make up the state vectors evolve according to the established physical laws (\textit{primitive concepts}), as in the real-world signals, see the biological signals, the theoretical and abstract state space becomes a space restricted to states with physical significance; it is from now on the so-called \textbf{phase space}. The graphic display of different trajectories traced by the state vectors in the phase space is termed phase diagram.

The graphical analysis of a dynamic system is reduced to a state space of at most three-dimensions, implicitly including the time variable\cite{Shelhammer2007}, without forgetting that other unknown variables, apart from the three considered, can be significant in the system dynamics. Since most real-world physical systems are nonlinear, with an ever-present coupled noise, from a graphic analysis, with the appropriate dynamic variables, at least the deterministic structure of the underlying system dynamics can be deducible, not apparent in a possible erratic evolution in the time domain. A more complex study, although less visual, with many more variables, requires a mathematical formalism not yet fully consolidated\cite{Cipra1993}.

In physiological terms, typically, in a clinical trial, it is customary to acquire a unique biological signal identifying each physiological response, so that not all the dynamic variables involved in the dynamics of the system under study are available. Thus, each biological signal represents a response of a physiological system; each response distinguishes the time evolution of a dynamic variable. Accordingly, from the scalar measurements that define the acquired biological signal, the different state vectors must be generated in order to recreate the different states that characterize the physiological system dynamics in question. The phase space reconstruction allows to faithfully reproduce the evolution of the system under study from a simple biological signal, in the absence of the dynamic variables involved.

\begin{figure}[ht!]
\centering
\subfloat[]{\includegraphics[width=0.47\columnwidth]{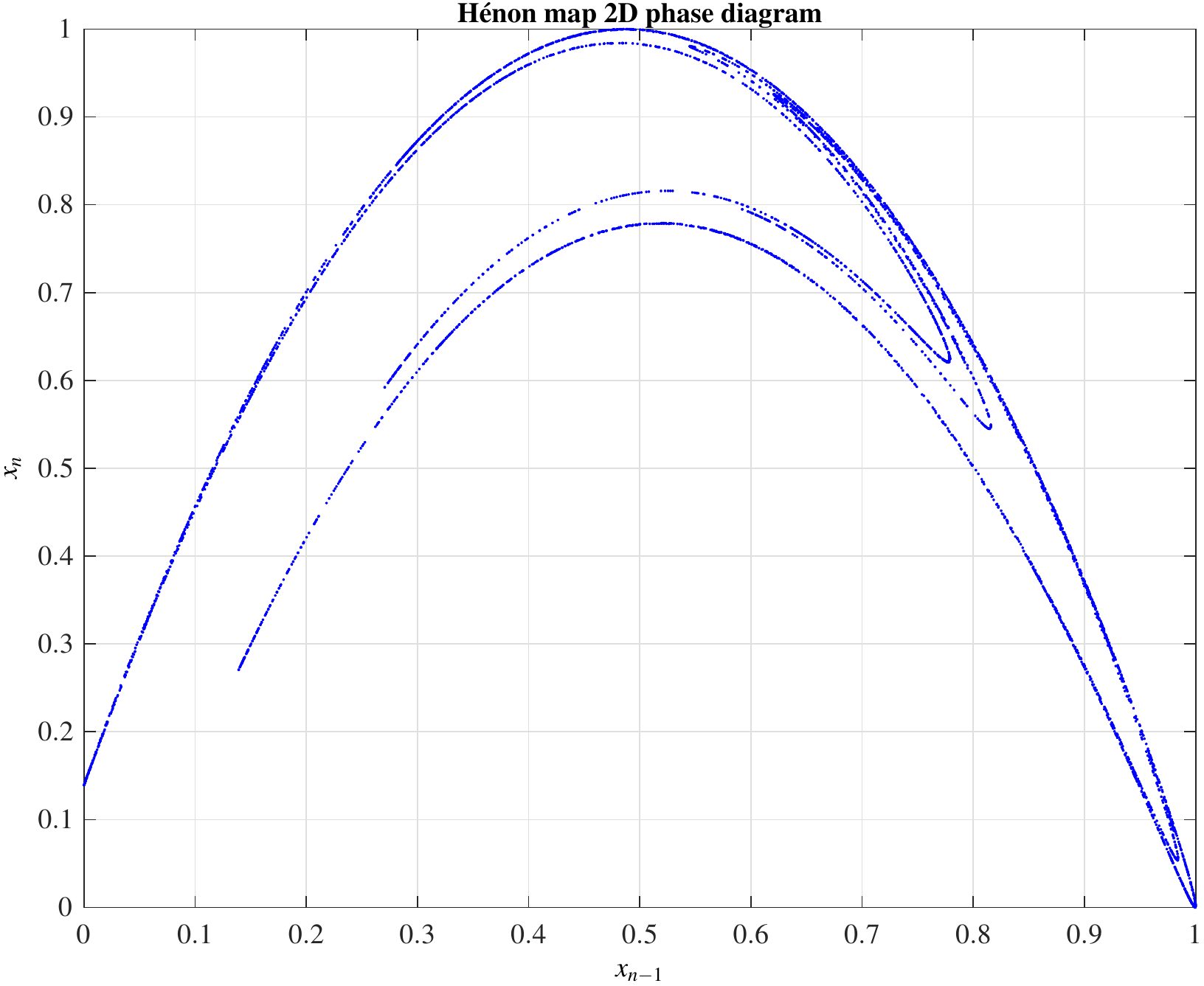}
\label{fig:F01a}}
\subfloat[]{\includegraphics[width=0.47\columnwidth]{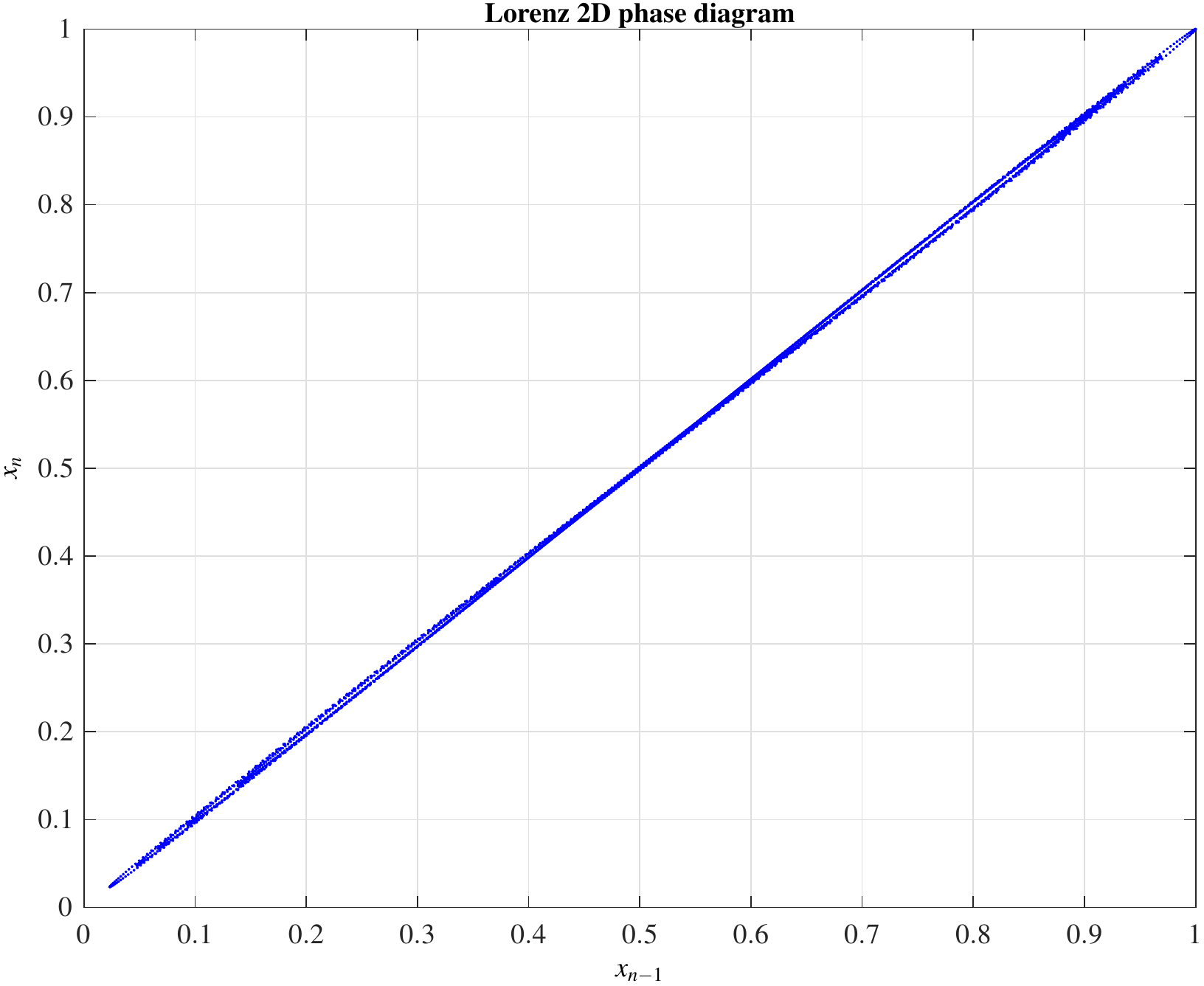}
\label{fig:F01b}}
\caption{2D phase diagram from the dynamic variable $x$ of the: (a) H\'enon map; (b) Lorenz flow, with sampling time of 2 ms.}
\label{fig:F01}
\end{figure}

Usually, observations or measures at regular time intervals, i.e., one every $\Delta t $  seconds, is an example of what in academic slang know as a time series. The most common method to define each state vector is to use delayed versions of the observations as state vectors components. In the simplest case, in a two-dimensional coordinate system, as illustrates Fig. \ref{fig:F01} for the two representative dynamical systems that we use in this paper to explain the methodology employed, the ordinates axis represents the measurement value $x_{n}$ at time $n\Delta t$ and the abscissas axis the measurement value $x_{n-1}$ at time $(n-1)\Delta t$. Hence, each point in the plot identifies the $\mathbf{x}_{n}=(x_{n-1},x_{n})$ state vector. 

At the most general level, the number $m$ of components of the state vector and the delay $\tau$ between components is variable, that is, $\mathbf{x}_{n}=(x_{n-(m-1)\tau},\dotsc,x_{n-\tau},x_{n})$. From a sequence of scalar measurements, it can reconstruct every state vector, in $m$-dimensional phase space, following the lag or delay method\cite{Liebert1989}. The state space reconstruction quality depends on the proper estimation of the parameters $\tau$ and $m$, considering the time delay embedding theorem\cite{Yule1927,Takens1981,Sauer1991}, which gives the ideal conditions under which the chaotic dynamical system can reconstruct from a sequence of observations of the state of the dynamical system.

The state or phase space reconstruction, in contrast to the traditional signal analysis in time or frequency domains, constitutes the cornerstone of the time series analysis in terms of nonlinear dynamics\cite{Eckmann1985, Broomhead1986}, and it is the first step in the analysis process of the real functional nature that reflects the physical system in question. In doing so, special attention must be given to determine the optimum values of $\tau$ and $m$\cite{Abarbanel1993} efficiently, so that the state space reconstruction faithfully allows to characterize the dynamics of the original system. We detail the most common techniques, elegant in its simplicity, which enable a first approximation of the state space reconstruction and then apply them to a particular biological signal, the Photoplethysmographic signal, extracting interesting conclusions that lead to promising future studies. 

In a typical trajectory, over the phase space, dynamic variables time evolution can tend to infinity or confined to a bounded region of the phase space. If the dynamic system is dissipative, once the transient response finishes, its dynamics will tend to a subset of the state space called \textit{attractor}. This subset is invariant under the dynamical system evolution. In a chaotic system, the attractors describe very complex geometric objects, having a typical fractal structure; they are the so-called \textit{strange attractors}.

The attractor geometry provides valuable information not only on the dynamic nature (Lyapunov exponents) of the underlying physical system but also about the structural complexity sustaining that dynamics (dimensionality), hence the vital importance of a successful phase space reconstruction. The connections between these factors go beyond the scope of this paper, though we will deal properly with these issues in future communications.

In this paper, we outline the due process to be followed in the phase space reconstruction from one scalar time series, applying the methodology to a relatively well-known and easily accessible biological signal, the Photoplethysmographic signal. In section \ref{sec:DET}, we introduce several graphical methods to intuitively assess the approximate determinism present in the dynamic system evolution, making the state space reconstruction meaningful. After, in the next section, in section \ref{sec:PHA}, we explain in general terms the main mechanisms underpinning the state space reconstruction, focusing mainly on how to properly determine the reconstruction parameters, the lag $\tau$ and the embedding dimension $m$. In section \ref{sec:BIO}, we briefly describe the basic characteristics of the biological signal used in this work, the PhotoPlethysmoGraphic (PPG) signal, and show the obtained results in line with the methodology referred to above. Finally, in section \ref{sec:CON}, we analyze and interpret the obtained results laying the ground for forthcoming works.

\section{Determinism and state space representation}\label{sec:DET}

From a linear perspective, the erratic (or irregular) behavior of the one system response is due to a random external component. The chaos theory finds that a random input is not the only cause to get out an irregular output. A simple deterministic equation, as it is with nonlinear chaotic systems, can generate irregular data without the contribution of any external input.

In a deterministic system, according to the deployed dynamical systems theory\cite{Hilborn2001}, once known its current state, the future states are entirely determined. The system dynamical and geometric properties all are to some extent included in the state space representation. Equations of motion that describe the behavior of a physical system as a function of time are deployed in the state space. Henceforth, the drawn trajectories in the state space represent the time evolution that the state vector is  undergoing over time. For this reason, the state space or phase space can be used to get an approach to the rules that govern the dynamical system evolution. 

\begin{figure}[ht]
\centering
\includegraphics[width=0.94\columnwidth]{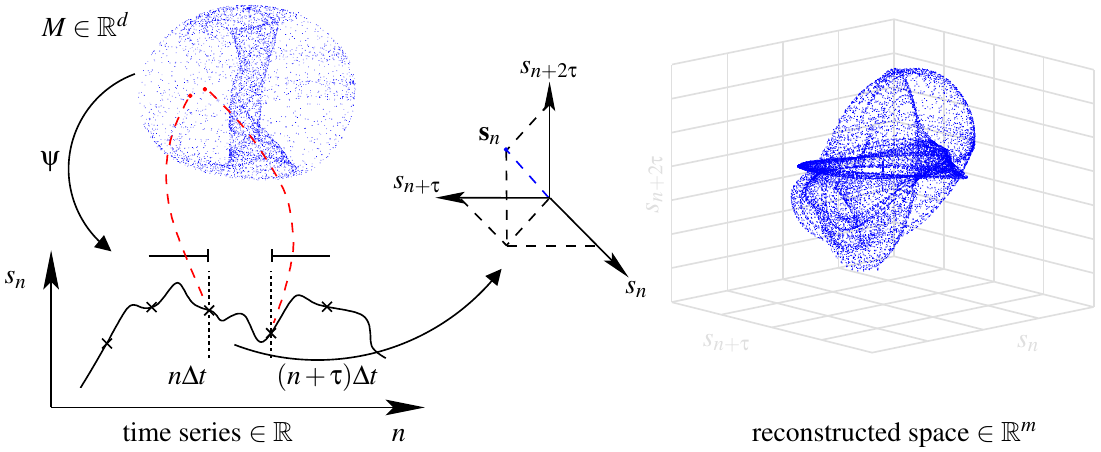}
\caption{State space reconstruction general overview.}\label{fig:F02}
\end{figure}

In an experimental setting, the system equations are not available. Usually, it is had solely with one system response projection in a one-dimensional space, a scalar time series, which represents the acquired measurements of the system in question. The state space reconstruction method aims to reconstruct the state vectors from the time series, so that the time evolution of these vectors replicates a dynamics equivalent to that of the original system, as shown in Fig. \ref{fig:F02}. A time series reflects roughly the one dynamical variable time evolution, but as J. Doyne Farmer once said the evolution of a variable depends on other variables of the system or, in other words, its value is within the history of other variables with which it interacts\cite{Gleick1988}. 

Ideally, a system of first-order ordinary differential equations taking action on the state space\cite{Eckmann1985, Strogatz1994, kantz2004} defines a dynamical system. A set of possibly infinite states, and certain transition rules that specify how the system moves from one state to another can describe several systems. For a finite-dimensional state space $\mathbb{R}^{m}$, an state is defined by a vector $\mathbf{x}\in \mathbb{R}^{m}$. Then the dynamics of the system can be described by an $m$-dimensional map or by a system of $m$ first-order ordinary differential equations called flow. In the first instance, such as the H\'enon map, the time is a discrete variable, 

\begin{equation}\label{eqn:E01}
\mathbf{x}_{n+1}=\mathbf{F}(\mathbf{x}_{n}),\quad n\in\mathbb{Z},
\end{equation}
while in the latter, as is the Lorenz flow, the time is a continuous variable, 

\begin{equation}\label{eqn:E02}
\frac{\text{d}}{\text{d}t}\text{\textbf{x}}(t)=\text{\textbf{f}}(\text{\textbf{x}}(t)),\quad t\in\mathbb{R}.
\end{equation}

For each initial condition $\text{\textbf{x}}_{0}$, or $\text{\textbf{x}}(0)$, the solutions to equations \eqref{eqn:E01} and \eqref{eqn:E02}, a sequence of points $\text{\textbf{x}}_{n}$, o $\text{\textbf{x}}(t)$, respectively, describe a dynamic system trajectory. With its time evolution a typical trajectory can tend to infinity or confine to a bounded region in the state space. All the initial conditions that lead to the same asymptotic behavior of the observed trajectory is known as \textit{basin of attraction}\cite{Ruelle1989}. 

The trajectory described in the state space, from a single observable, is a presumable indication of the presence of a deterministic behavior if the states not arrange as a point cloud, despite an erratic appearance in the time domain. In a 2D phase diagram the variable time, on the abscissa axis, is replaced by the observable value in a prior time determined by the parameter $\tau$, as discussed in the section \ref{sec:IN}; in a 3D phase diagram except for the first axis, all other axes pick up previous values of the observable variable separated from each other a time $\tau\Delta t$. 

\subsection{Poincar\'e section}\label{ssec:POSE}

\begin{figure}[ht!]
\centering
\subfloat[]{\includegraphics[width=0.47\columnwidth]{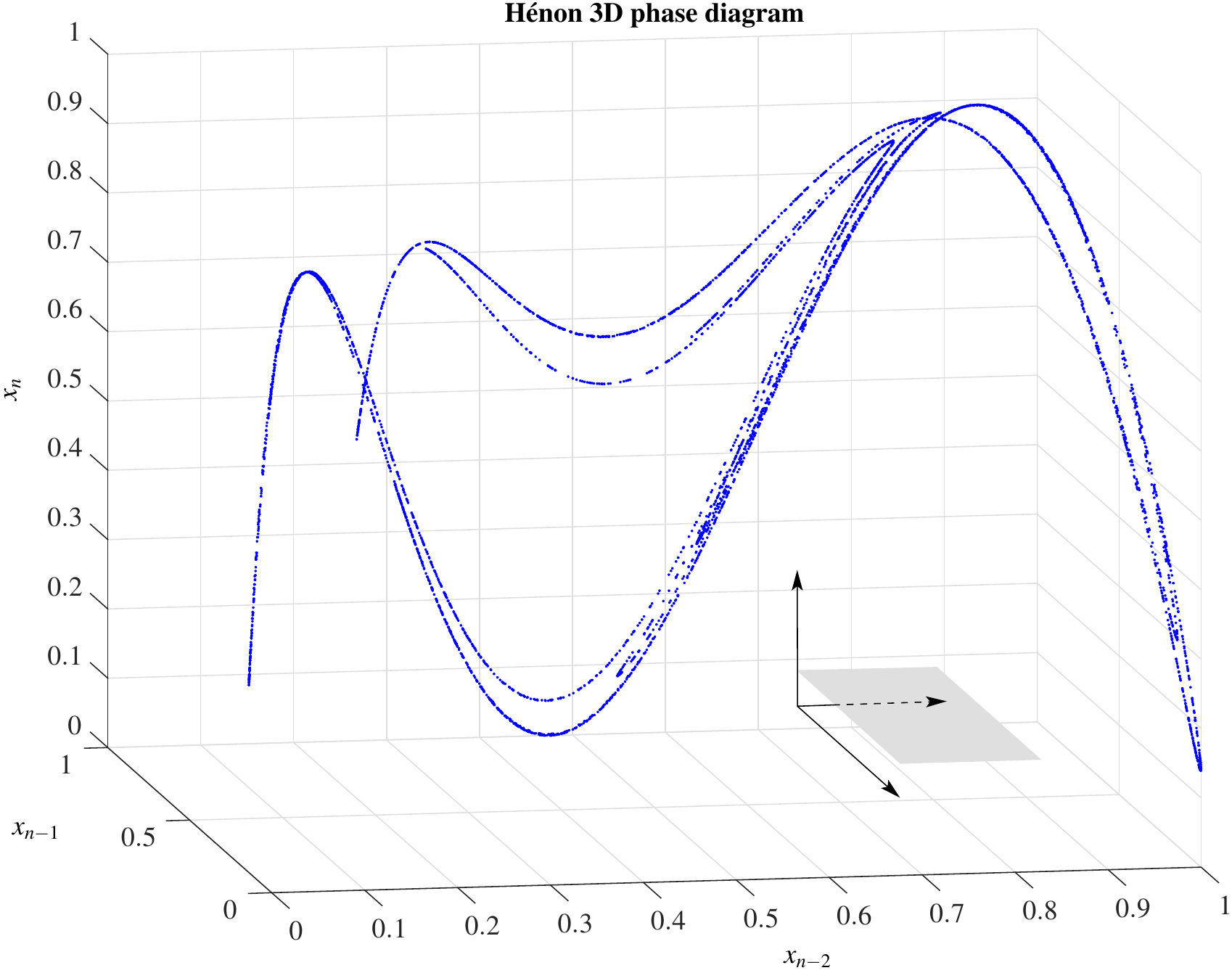}
\label{fig:F03a}}
\subfloat[]{\includegraphics[width=0.47\columnwidth]{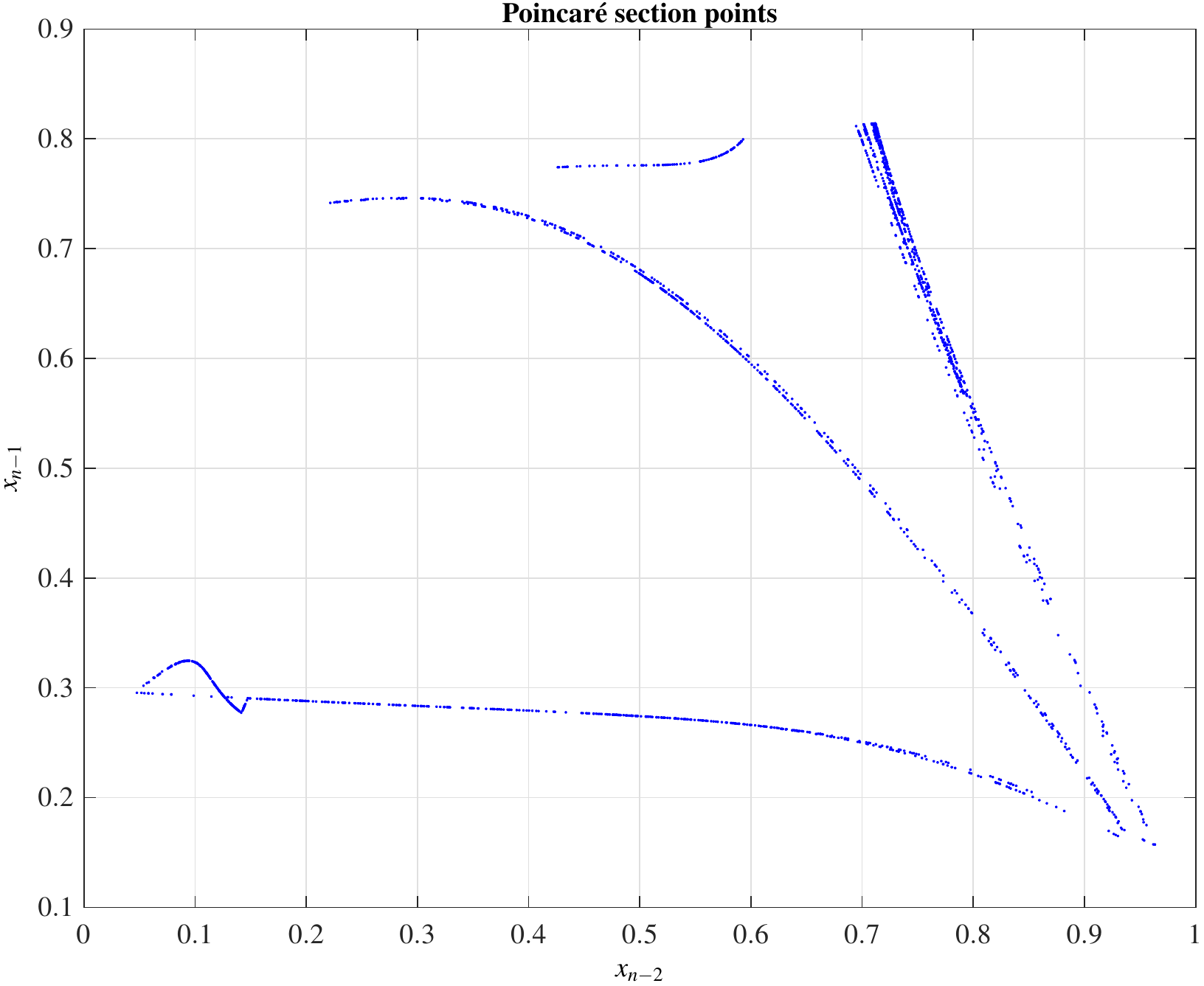}
\label{fig:F03b}}
\caption{From the dynamic variable $x$ of the H\'enon map: (a) 3D phase diagram with $\tau=1$ and Poincar\'e section's orientation (inner sketch); (b) points collected in the Poincaré section.}
\label{fig:F03}
\end{figure}

It covers the geometric figure that describes the evolution of a trajectory in a cross-section of an attractor, transversal to the flow or bundle of trajectories, as highlighted in the inner plane in Fig. \ref{fig:F03a}, gray coloured. Fig. \ref{fig:F03b} shows the points contained in the Poincar\'e section.

The Poincar\'e section aims at facilitating a complementary perspective of the general dynamics of the system (internal structure of the attractor), contributing to identify the type of attractor\cite{Wiggins2003}. The Poincar\'e section describes a pattern of points that can be labeled chronologically, although the time between successive intersections with the section is irrelevant. It must not be confused with a Poincar\'e map, also called a \textit{return map}, which reconstructs the temporal sequence of spatially arranged points on the Poincar\'e section. In a chaotic attractor, the Poincar\'e section usually describes a complex geometry, as reflected in Fig. \ref{fig:F03b}, with a very fine distinctive structure, adopting shapes with texture or with multiple layers (fractals).

\subsection{Other plots}\label{ssec:OTPL}

The following items summarize other methods of visualizing a possible hidden determinism. 

\begin{figure}[ht!]
\centering
\subfloat[]{\includegraphics[width=0.47\columnwidth]{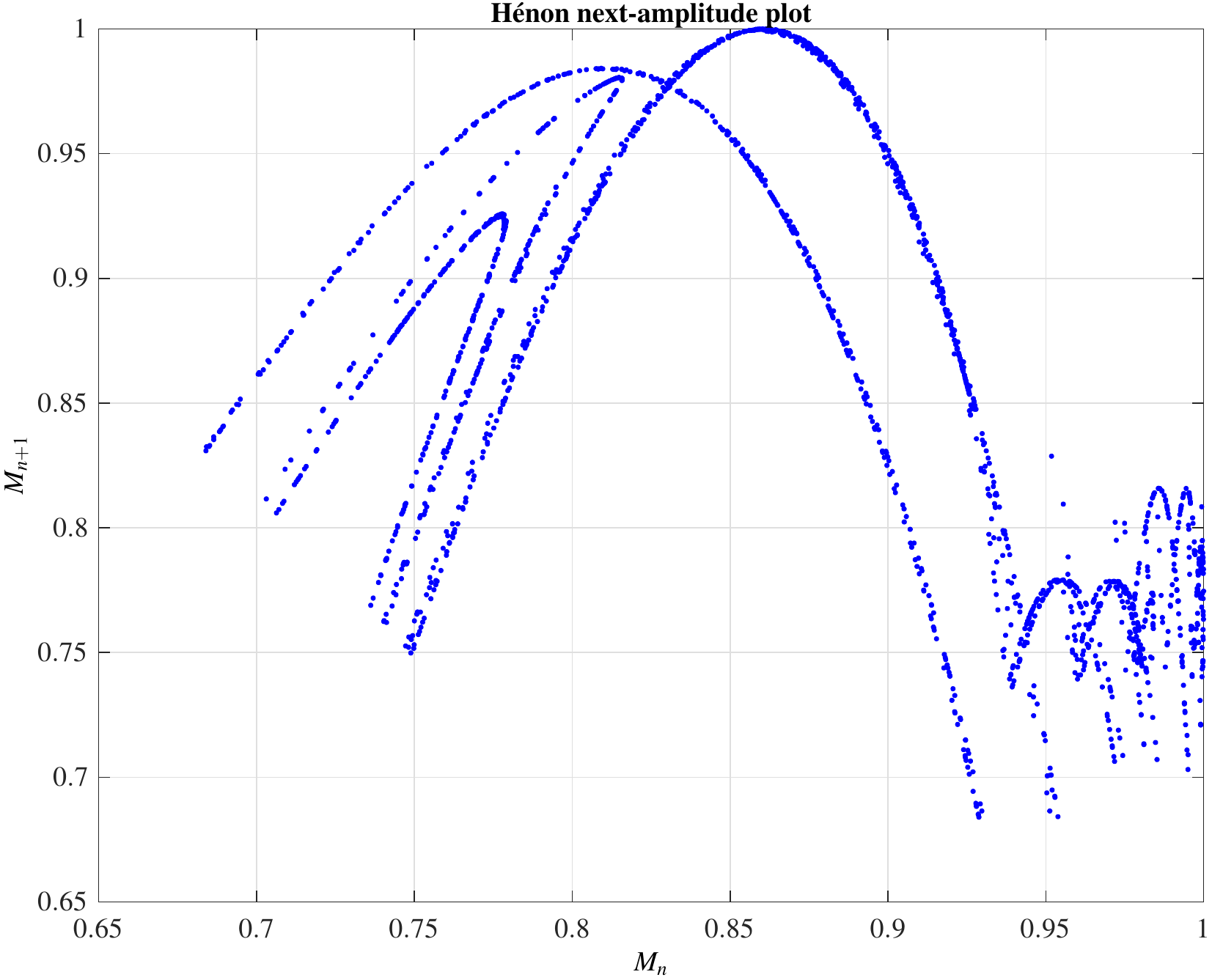}
\label{fig:F04a}}
\subfloat[]{\includegraphics[width=0.47\columnwidth]{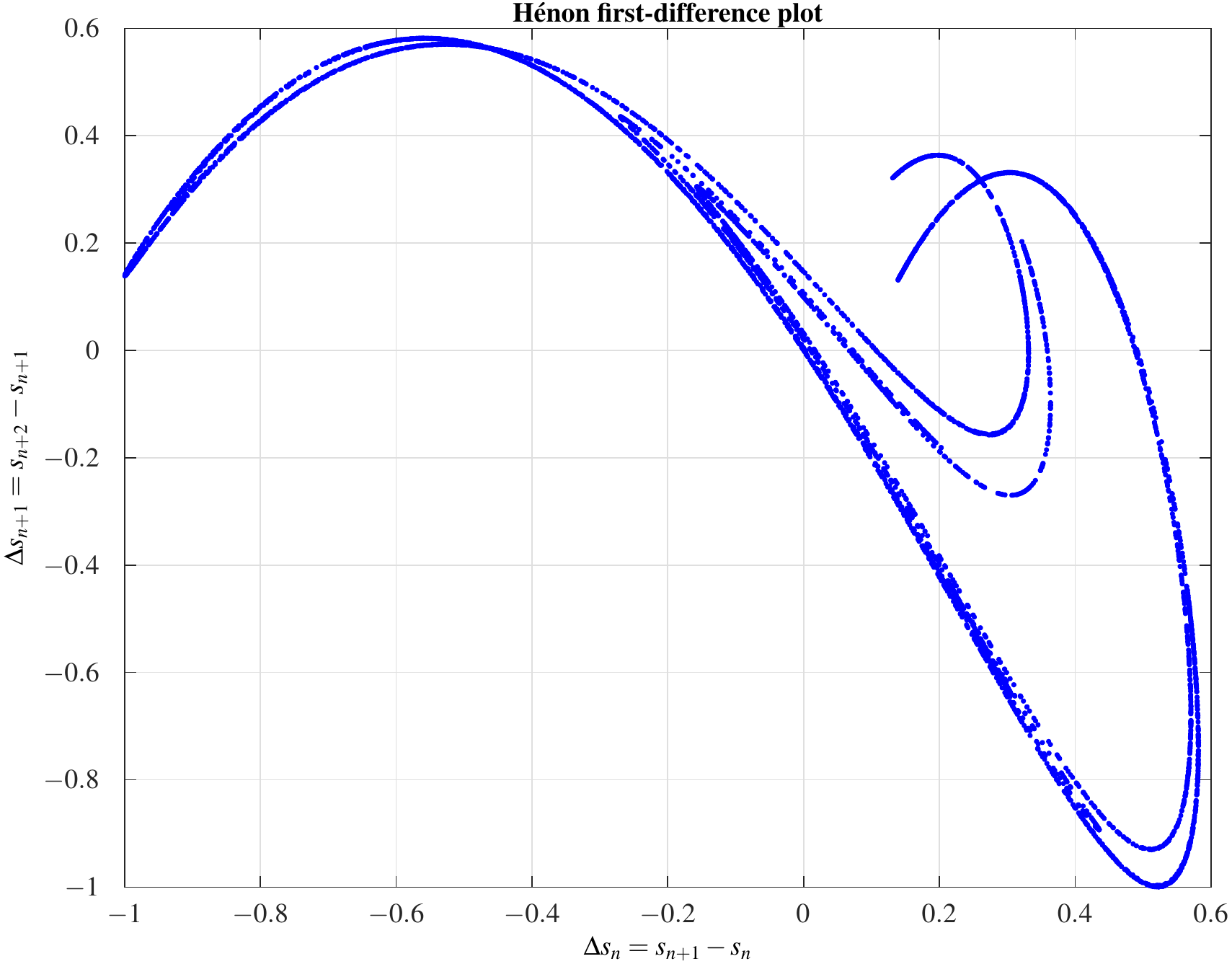}
\label{fig:F04b}}\hfil
\subfloat[]{\includegraphics[width=0.47\columnwidth]{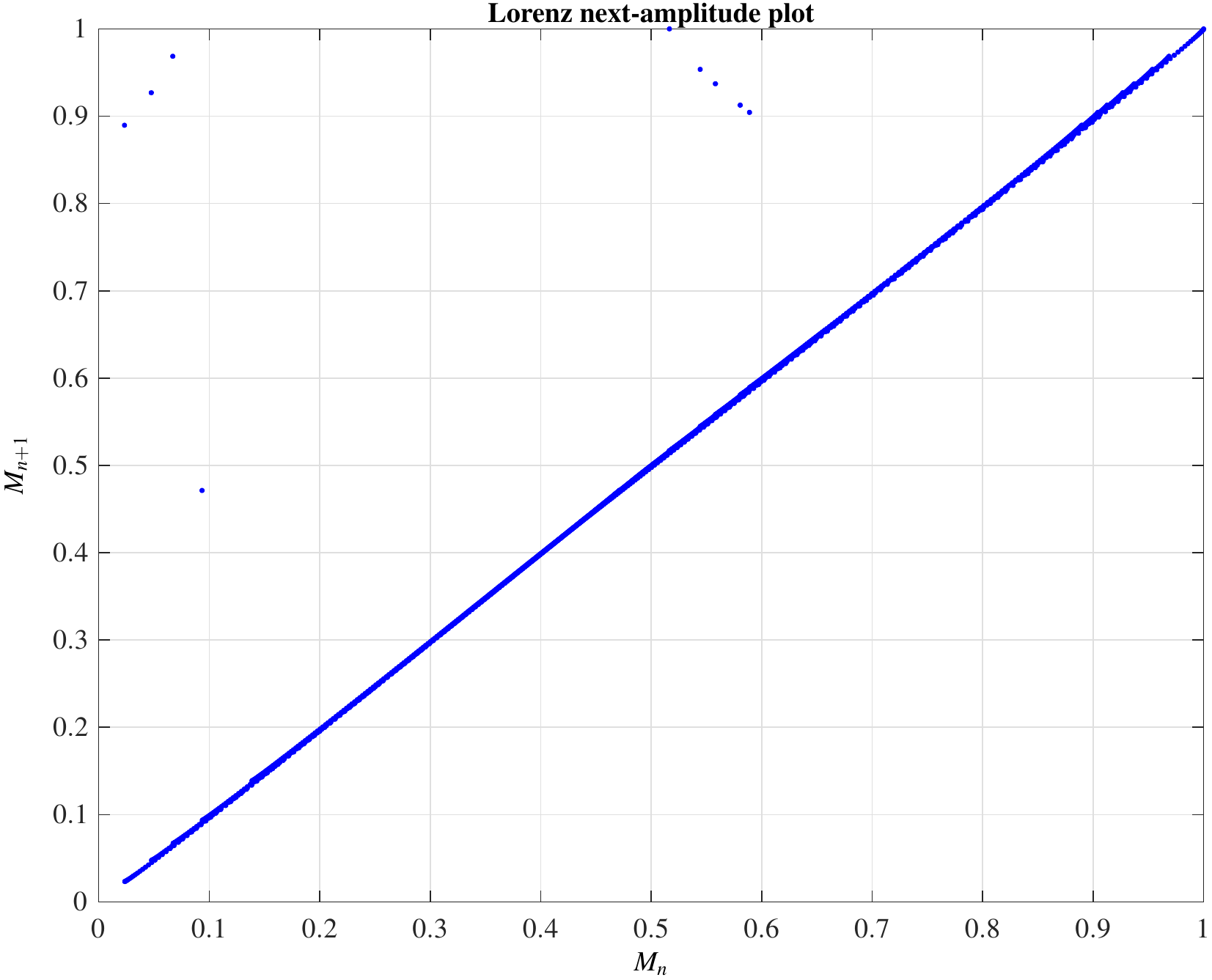}
\label{fig:F04c}}
\subfloat[]{\includegraphics[width=0.47\columnwidth]{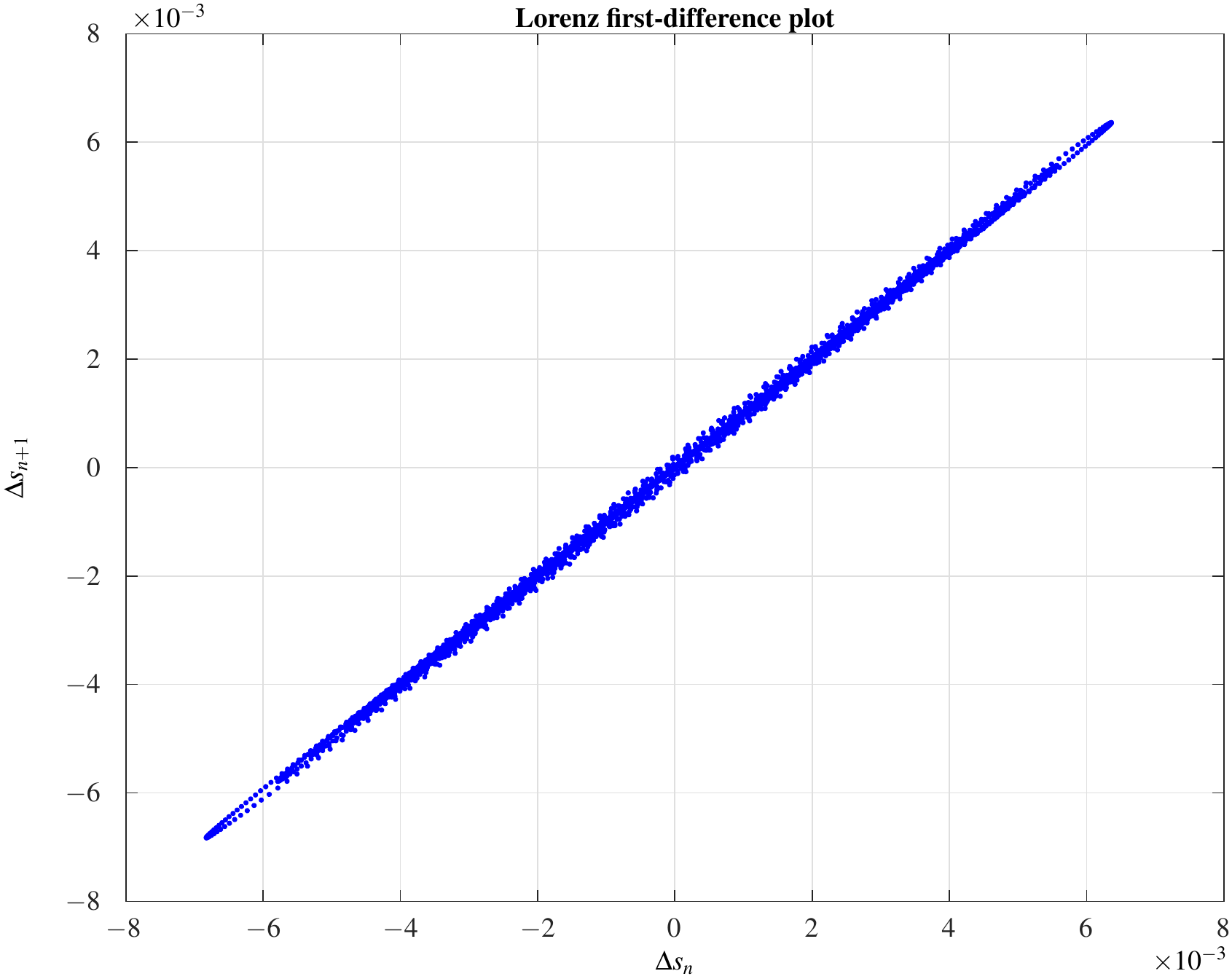}
\label{fig:F04d}}\hfil
\subfloat[]{\includegraphics[width=0.47\columnwidth]{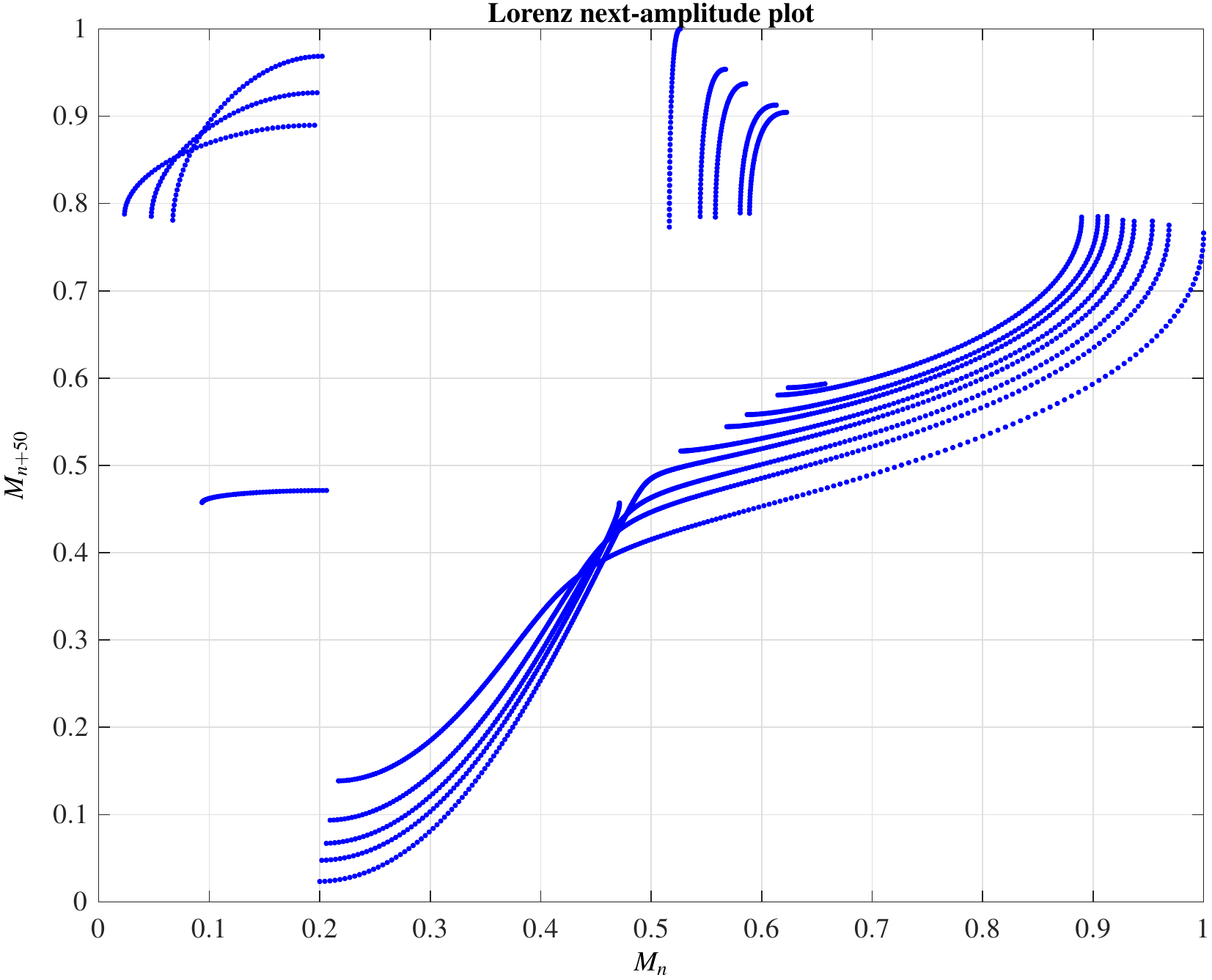}
\label{fig:F04e}}
\subfloat[]{\includegraphics[width=0.47\columnwidth]{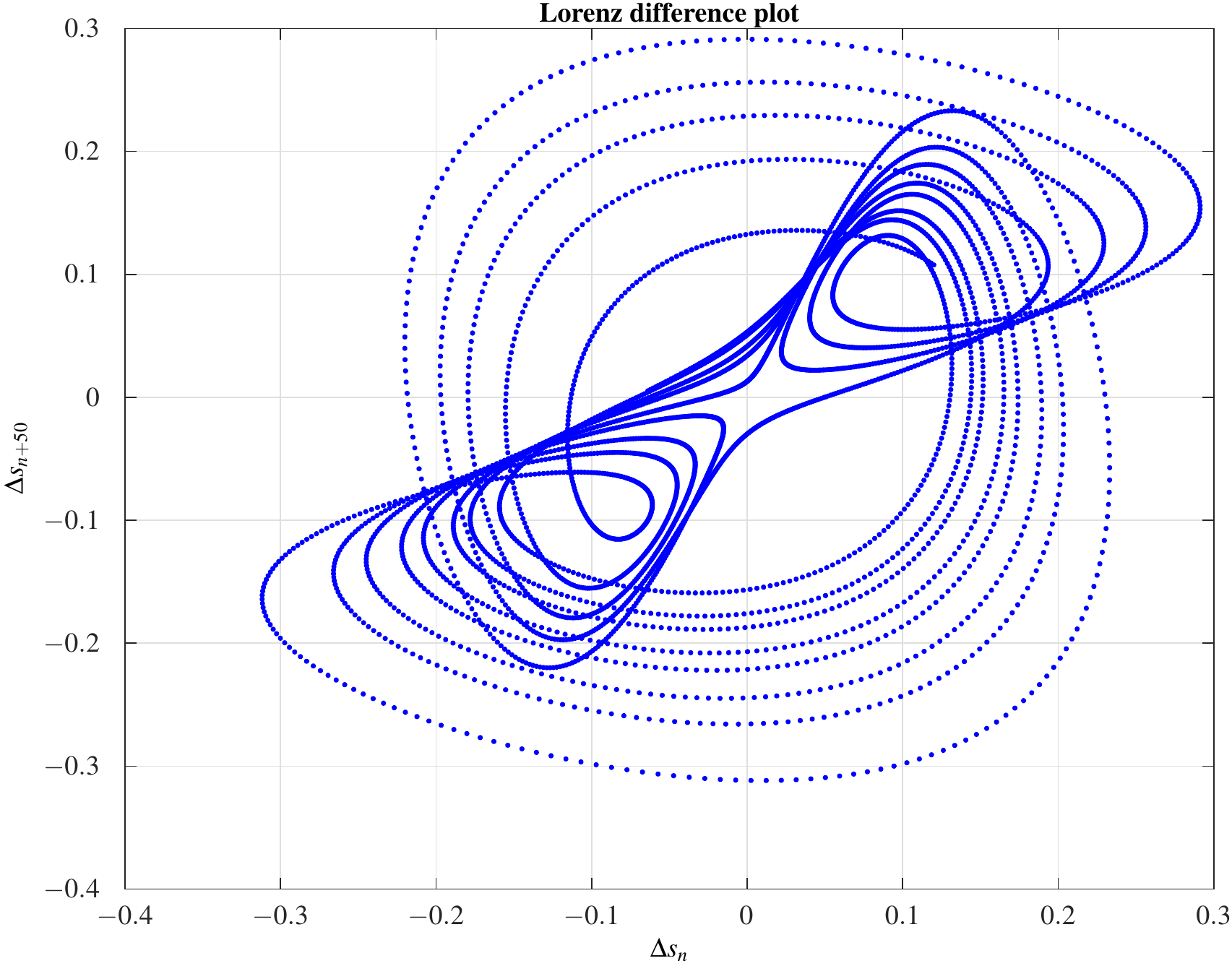}
\label{fig:F04f}}
\caption{Next-amplitude and difference plots, from the dynamic variable $x$ of the H\'enon map and the Lorenz flow with sampling time of 2 ms: (a)-(b) H\'enon map with $\tau=1$; (c)-(d) Lorenz flow with $\tau=1$; (e)-(f) Lorenz flow with $\tau=50$.}
\label{fig:F04}
\end{figure}

\begin{itemize}
\item[a.] \textit{Next-amplitude plot}. From successive maximums detected in time series, each one $M_{n}$ is represented (abscissa) versus its immediate or subsequent successor $\tau$ times $M_{n+\tau}$ (ordinate) ahead. A well-defined curve, like the one in Fig. \ref{fig:F04a}, could reveal the presence of chaos, although the noise could mask a correct interpretation.
\item[b.] \textit{Difference plot}. The graphic's coordinates are delayed differences between successive observations, whether immediate or separated $\tau$ number of times. On the abscissa axis it is represented $\Delta s_{n}=s_{n+1}-s_{n}$ and on the ordinate axis the next difference $\Delta s_{n+\tau}=s_{n+\tau+1}-s_{n+\tau}$. In the simplest form, the \textit{first-difference plot}, with a delay $\tau=1$, on the abscissa it is represented $s_{n+1}-s_{n}$ and on the ordinate $s_{n+2}-s_{n+1}$. The presence of an infinitely continuous curve, as illustrated in Fig. \ref{fig:F04b}, evidences a high degree of underlying determinism.
\end{itemize}

\section{Phase space reconstruction method}\label{sec:PHA}

It was in 1980 that Packard \textit{et al}.\cite{Packard1980} first tackled the problem of how to connect the phase space or state space vector $\mathbf{x}(t)$ of the dynamical variables of one physical system to a possible time series $\{s_{n}\}$ measured in any experiment. 

A time series is a sequence of scalar measures $s_{n}$ of an observable, acquired at regular times $\Delta t$. The different measures depending on the current state of the system, 

\begin{equation}\label{eqn:E03}
s_{n}=\psi(\text{\textbf{x}}(n\Delta t))+\eta_{n},
\end{equation}
where $\psi$ represents an observable measurement function and $\eta_{n}$ the \textit{measurement noise} which characterizes the random nature of the imprecision of the measure\cite{Casdagli1991}.

An $m$-dimensional reconstruction of the $\mathbf{s}_{n}$ state vectors, from scalar measurements, is given by

\begin{equation}\label{eqn:E04}
\text{\textbf{s}}_{n}=(s_{n-(m-1)\tau},\dotsc,s_{n-\tau},s_{n}),\quad \tau\in\mathbb{Z}^{+}.
\end{equation}

The time interval between adjacent coordinates of the state vector is $\tau\Delta t$, and it is known as lag or delay time. Formally It has been shown that if $m$ is higher than two times the capacity dimension of the attractor ($\text{dim}_{\text{C}}$), a one-to-one correspondence between the reconstructed attractor and the real attractor is guaranteed, no matter how large the dimension of the original state space is\cite{Takens1981,Sauer1991}.

\subsection{Lag or delay time selection}\label{ssec:LAG}

\begin{figure}[ht!]
\centering
\subfloat[]{\includegraphics[width=0.47\columnwidth]{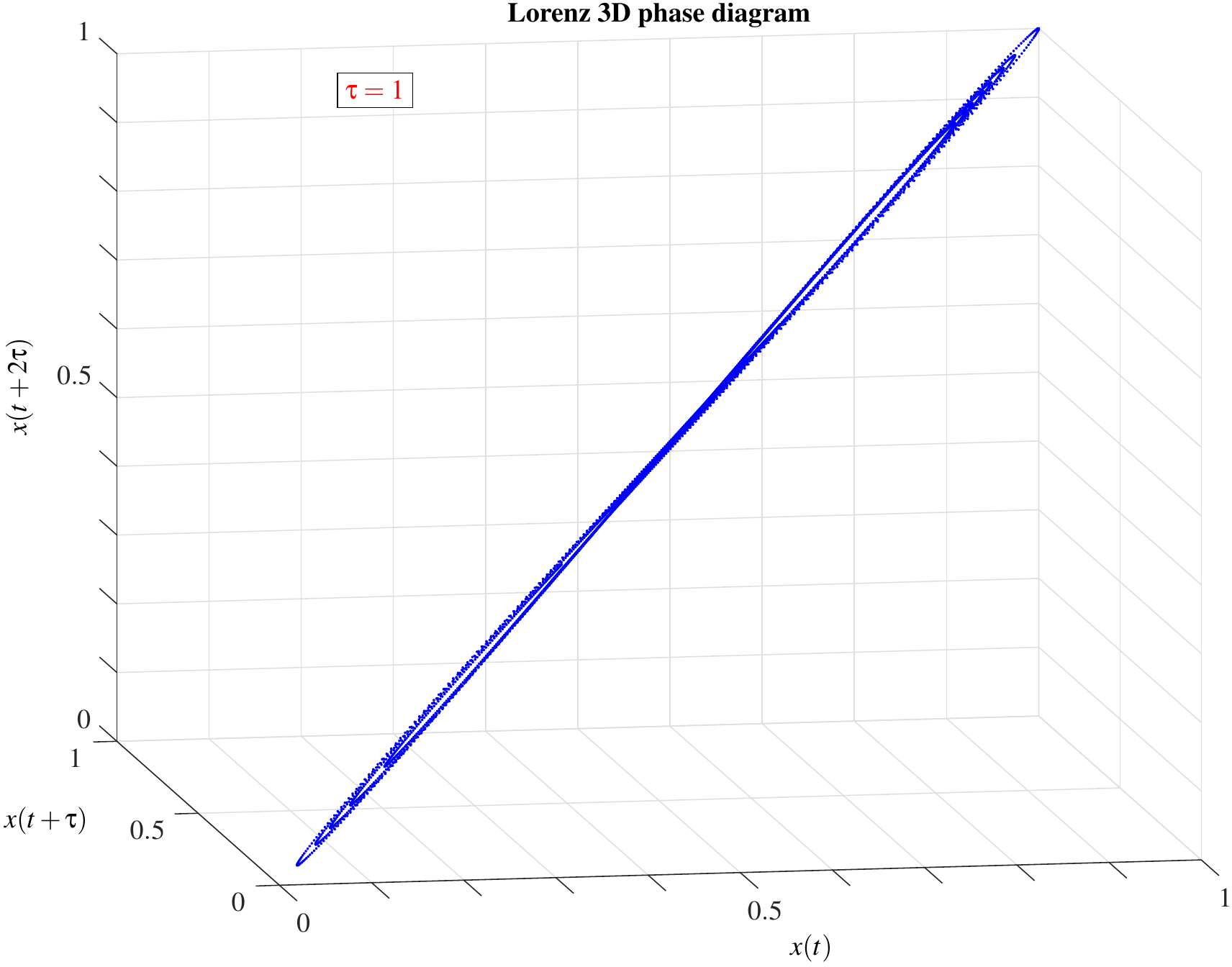}
\label{fig:F05a}}
\subfloat[]{\includegraphics[width=0.47\columnwidth]{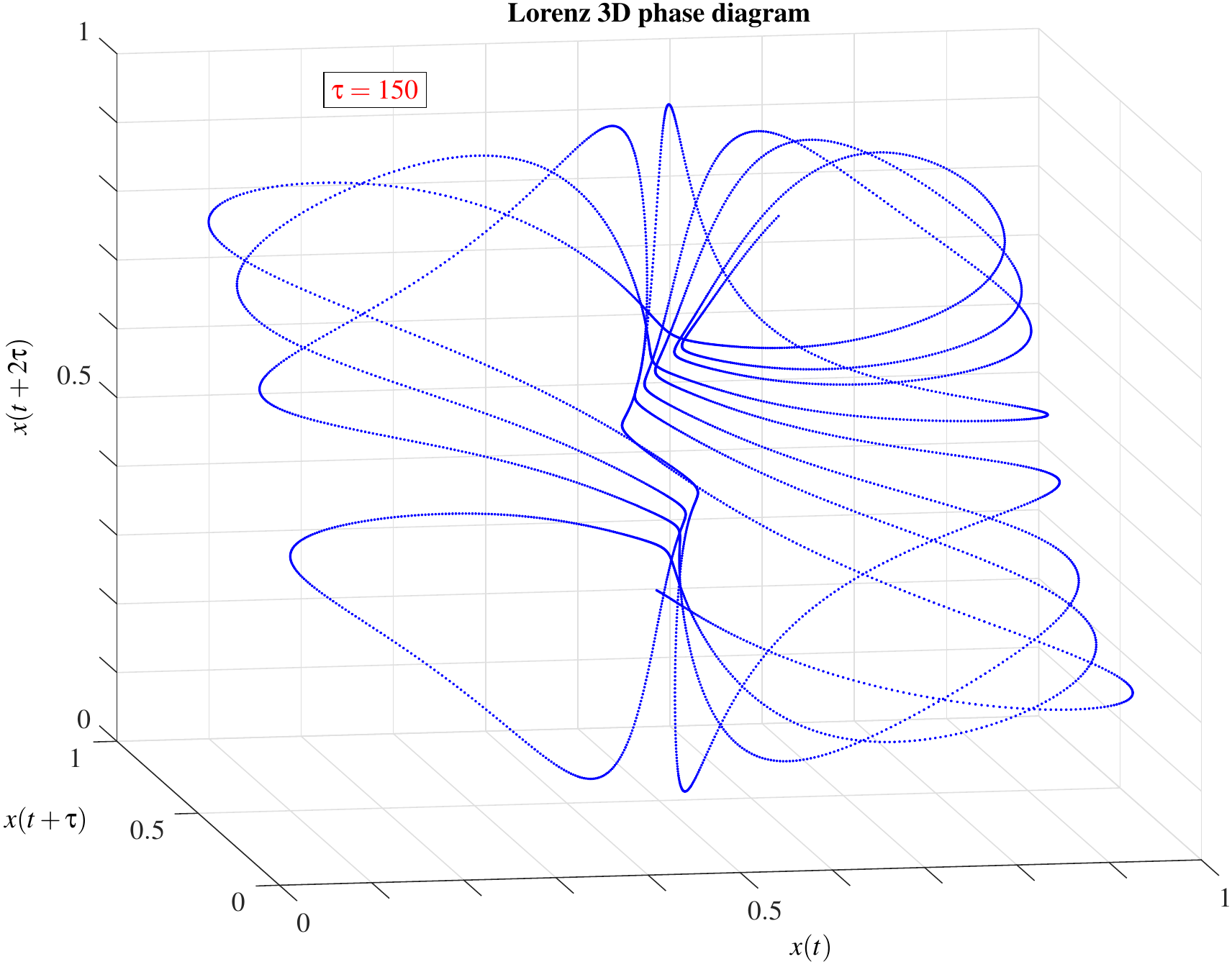}
\label{fig:F05b}}
\caption{State space reconstruction from the dynamic variable $x$ of
the Lorenz attractor, with a sampling time $\Delta t$ of 2 ms: (a) with $\tau=1$; and
(b) with $\tau=150$.}
\label{fig:F05}
\end{figure}

In theory, with an unlimited number of measurements without noise, any delay is equally valid. In practice, with real experimental data, the geometrical and dynamic properties that characterize the attractors differ from each other, depending on the delay value chosen\cite{kantz2004}. If the sampling time $\Delta t$ is tiny, the values of consecutive samples are very similar, $s_{n}\approx s_{n+\tau}$, and with $\tau=1$ the represented points contain much redundant information. Also, with a small delay, the noise level can blur or hide any local geometric structure. Regardless of the form that the attractor can take, the graphic arrangement of perfectly related pairs of points describes a straight line, as an identity function (a diagonal line of $45^{\circ}$) without any meaning, as illustrated in Fig. \ref{fig:F05a}. 

In order for the distribution of points to provide some geometric significance, the values of the components that make up the coordinate system must be sufficiently independent. In the real state space, in which the coordinates correspond to different variables, this condition is satisfied. 

In the same vein, a delay too large can be counterproductive. With a significant delay the dynamic relationship between the values of the variable disappears, so that, as the delay value increases, the geometric structure of the points becomes more complex and diffuse until finally the points are dispersed randomly on the state space (see Fig. \ref{fig:F05b}).

\begin{figure}[ht]
\centering
\includegraphics[width=0.94\columnwidth]{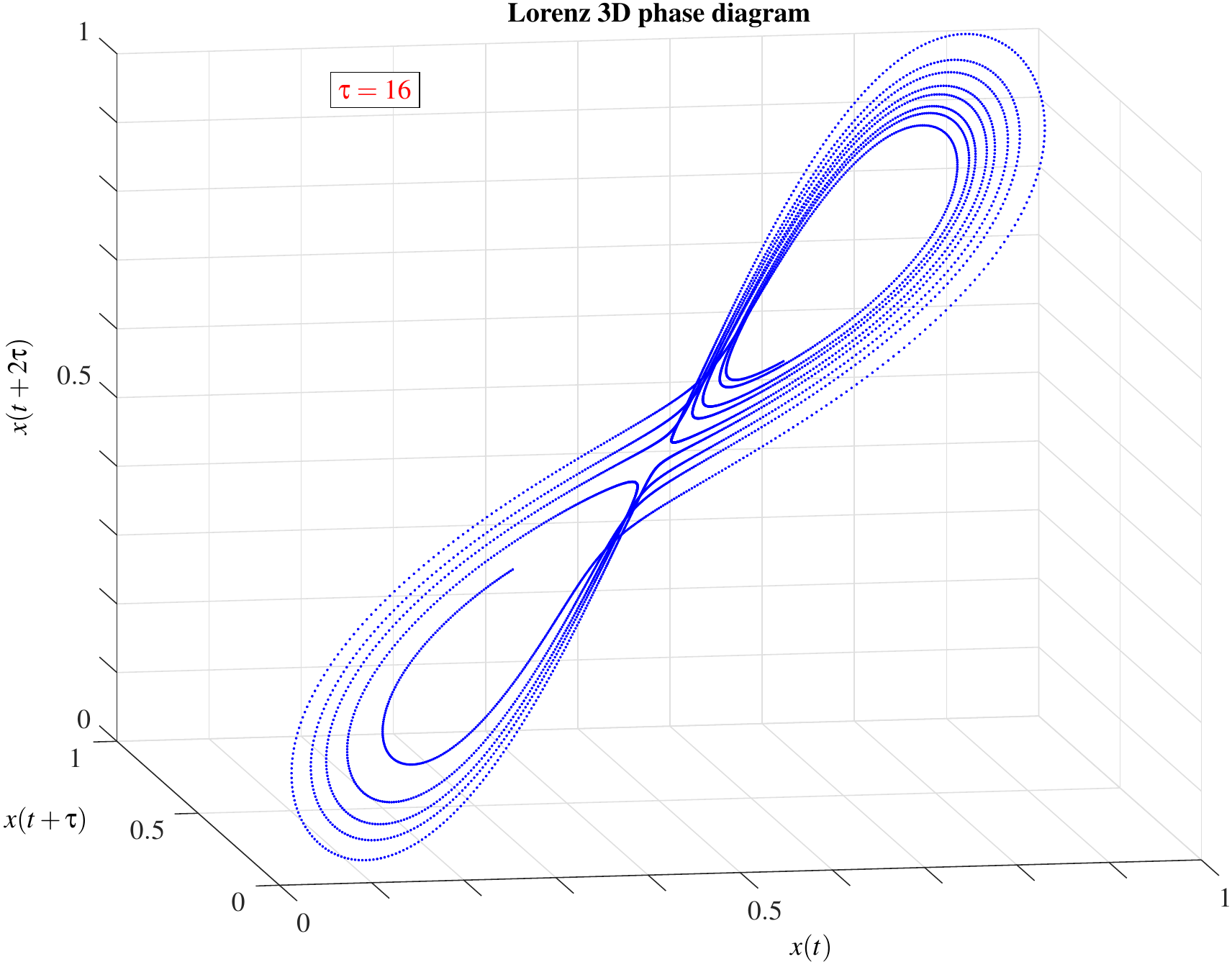}
\caption{State space reconstruction from the dynamic variable $x$ of
the Lorenz attractor, with a sampling time $\Delta t$ of 2 ms and $\tau=16$.}\label{fig:F06}
\end{figure}

In Fig. \ref{fig:F06} the Lorenz flow, with the correct delay time $\tau=16$, is represented. Next subsections describe how to calculate it.

\subsubsection{Autocorrelation coefficients}

The autocorrelation coefficients $R_{\tau}$ measure the correlation degree of a variable with itself at different instances. Its constituents are autocovariance and variance.

\begin{eqnarray}\label{eqn:E05}
R_{\tau} &= & \frac{\text{autocovariance}}{\text{variance}} \nonumber \\ 
& = & \frac{\frac{1}{N}\sum_{i=1}^{N-\tau}\left(s_{i}-\left\langle s\right\rangle\right)\left(s_{i+\tau}-\left\langle s\right\rangle\right)}{\frac{1}{N}\sum_{i=1}^{N}\left(s_{i}-\left\langle s\right\rangle\right)^2},
\end{eqnarray}
where $N$ represents the time series length, and $\left\langle s\right\rangle$ the arithmetic mean of all the time series observations. 

According to the equation \eqref{eqn:E05}, with a minimum delay $\tau=0$, the coordinates of each point are identical ($s_{n}=s_{n+\tau}$), and the autocorrelation is maximum $R_{\tau=0}=1$. As delay increases, autocorrelation decreases until it eventually is reduced to zero, or, as happens in real data, there is a fluctuation around zero, within a narrow margin, due to the noise in the data. In short, the autocorrelation coefficients range from $+1$ y $-1$. Maximum values, $R_{\tau}=\pm 1$, are perfectly correlated data; minimum values, $R_{\tau}=0$, correspond to uncorrelated data.

Autocorrelation coefficients, for successive delays, conform the autocorrelation function. It is also known as the spectral autocorrelation coefficient\cite{Davis2011} (it is recommended to use $\left\lfloor N/4\right\rfloor$ delays with a time series of $N>50$ observations). The graphical representation of $R_{\tau}$ is known as correlogram. To a certain extent, a correlogram shows the type of regularity in the data. In the case of trendless and uncorrelated data, 95\% of the autocorrelation coefficients are contained, in theory, in a band of $ \pm 2/\sqrt {N}$, around zero. About 5\% of them can exceed the indicated limit without lost its condition of uncorrelated data\cite{Makridakis2008}.

\begin{figure}[ht!]
\centering
\subfloat[]{\includegraphics[width=0.47\columnwidth]{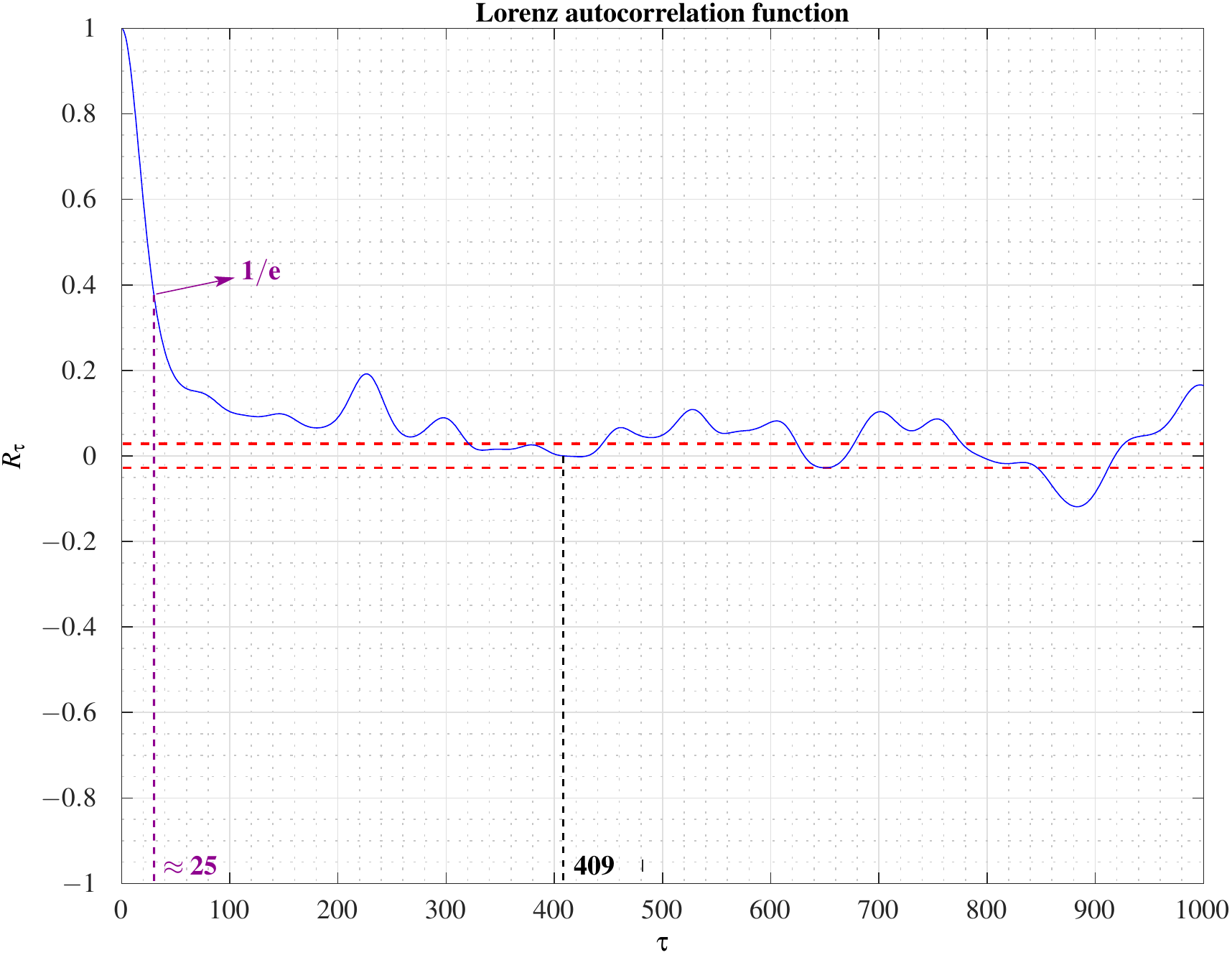}
\label{fig:F07a}}
\subfloat[]{\includegraphics[width=0.47\columnwidth]{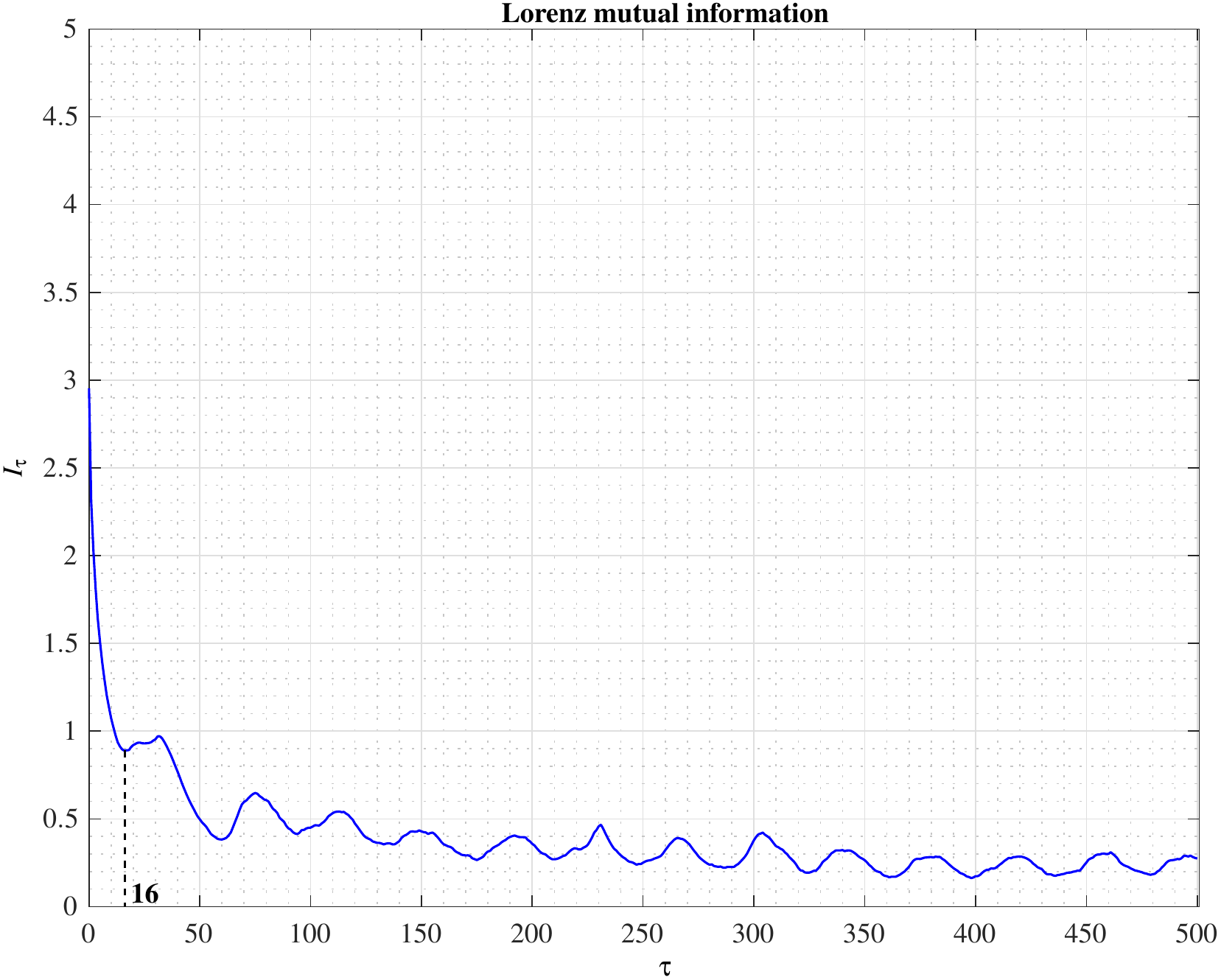}
\label{fig:F07b}}
\caption{From the dynamic variable $x$ of the Lorenz attractor: (a) Autocorrelation Function (AF), where the first zero crossing is when $\tau=409$; (b) Mutual Information (MI), where the first minimum is with $\tau=16$.}
\label{fig:F07}
\end{figure}

From the autocorrelation function, it is possible to adopt some criteria for selecting the optimal delay. Among the various selection criteria, the most commonly used determines as optimal delay the first zero crossing of the autocorrelation function, as illustrated in Fig. \ref{fig:F07a}. As an example of an alternative approach,  in climate-related attractors\cite{Tsonis2012}, other criteria apply the one known as correlation time (when $R_{\tau}$ decreases to a value of  $1/e\approx 0.37$).

\subsubsection{Mutual information}

Autocorrelation coefficients consider the degree of the mutual relationship on a linear basis, inappropriate in nonlinear systems\cite{Abarbanel1993}. In these situations, it is best to use the mutual information, as it enables to determine, on a probabilistic basis, to what extent two values of the same variable, measured at different time instants, relate to each other. For example, if the point coordinates are identical, with $\tau=0$, these represent the same information and, therefore, one coordinate accurately predicts the other, or, otherwise, the amount of information that one coordinate contains about the other, mutual information, is maximal.

Whenever there is any relationship between two values of the same variable, acquired at different time instants, one value contains information about the other value. That is, one value helps predict the other value or, in other words, the knowledge of one value reduces the uncertainty of the other value. The uncertainty reduction is called mutual information. If $\mathbf{X}$ denotes a time series, $\{s_{n}\}$, ergo, $s_{1},s_{2},\dotsc,s_{N}$, with $N$ observations, and $\mathbf{Y}$ the delayed time series, $\{s_{n+\tau}\}$, ergo, $s_{1+\tau},s_{2+\tau},\dotsc,s_{N}$, con $N-\tau$ observations, where $\tau$ indicates the delay, the average mutual information $I_{\mathbf{Y};\mathbf{X}}$ between both time series can be expressed, in probabilistic terms, as

\begin{eqnarray}\label{eqn:E06}
I_{\mathbf{Y};\mathbf{X}}=I_{\tau}&=&\sum_{i=1}^{N_{c}}\sum_{j=1}^{N_{c}}P(x_{i},y_{j})\log\frac{P(x_{i},y_{j})}{P(x_{i})P(y_{j})}\nonumber \\
 &=& \sum_{i=1}^{N_{r}}P(s_{i},s_{i+\tau})\log\frac{P(s_{i},s_{i+\tau})}{P(s_{i})P(s_{i+\tau})},
\end{eqnarray}
where $N_{c}$ is the number of cells containing points, with non-zero probability, and $N_{r}$ is the number of routes, $s_{i}s_{i+\tau}$, in the state space. Equation \eqref{eqn:E06}, in terms of entropy, rewrites as

\begin{equation}\label{eqn:E07}
I_{\mathbf{Y};\mathbf{X}}=H_{\mathbf{Y}}+H_{\mathbf{X}}-H_{\mathbf{X},\mathbf{Y}},
\end{equation}
where $H_{\mathbf{X}}$ is the entropy of $\mathbf{X}$, $H_{\mathbf{Y}}$ the entropy of $\mathbf{Y}$, and $H_{\mathbf{X},\mathbf{Y}}$ the joint entropy of $\mathbf{X}$ and $\mathbf{Y}$. So somehow, the mutual information involves a measure of predictability of the system, that is, a measure of the degree of knowledge of $s_{n+\tau}$ noted $s_{n}$.

In order to reconstruct an attractor, as faithfully as possible, minimum mutual information and delay value are required. As with the autocorrelation function, the mutual information decreases, as the delay increases, until it eventually becomes zero. The first minimum of the mutual information includes a possible criterion for selecting the optima delay value\cite{Liebert1989}, as shown in Fig. \ref{fig:F07b}. One limitation of this method is that many points are needed if we are to get a consistent result.

\subsection{Embedding dimension selection}\label{ssec:EMB}

There is no rule of thumb to set the minimum reconstruction dimension $m$, and none of the published proposals is widely accepted, so it is a good idea to use more than one method on the same data. Among all possible techniques, the correlation dimension and \textit{false nearest neighbors} stand out, though principal components analysis, as a preliminary attempt, may shed some light.

From a scalar time series, $\{s_{n}\}_{n = 1}^{N}$ acquired, and with the proposed $\tau$ value as stated above, $N-(m-1)\tau$ vectors with $m$ component per vector are defined, where each component symbolizes an alleged dynamic variable of the physical system. To a certain extent, we initiate a multivariate analysis with $m$ time series data obtained by \eqref{eqn:E04}.

\begin{figure*}[ht!]
\centering
\subfloat[]{\includegraphics[scale=0.32]{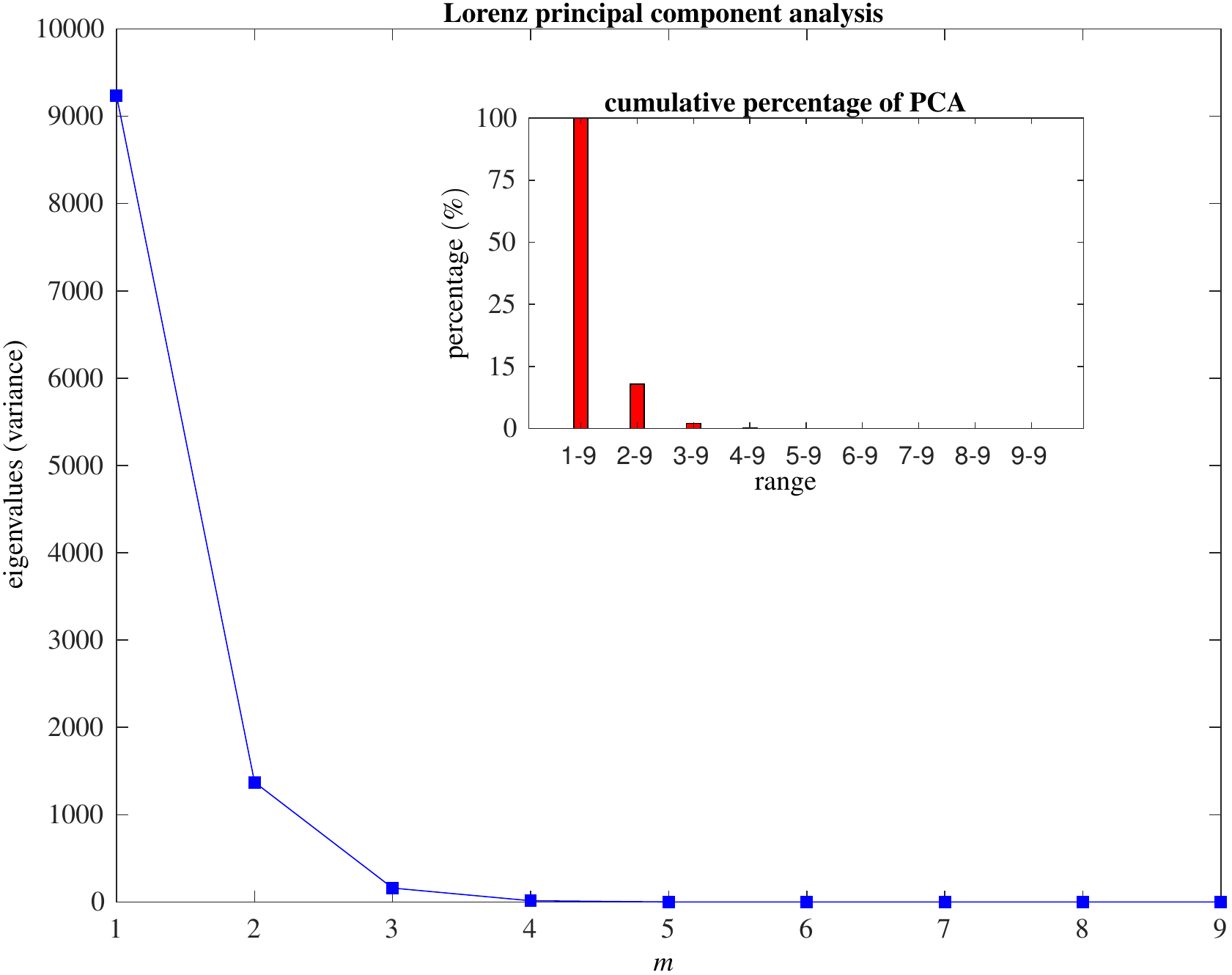}
\label{fig:F08a}}
\subfloat[]{\includegraphics[scale=0.32]{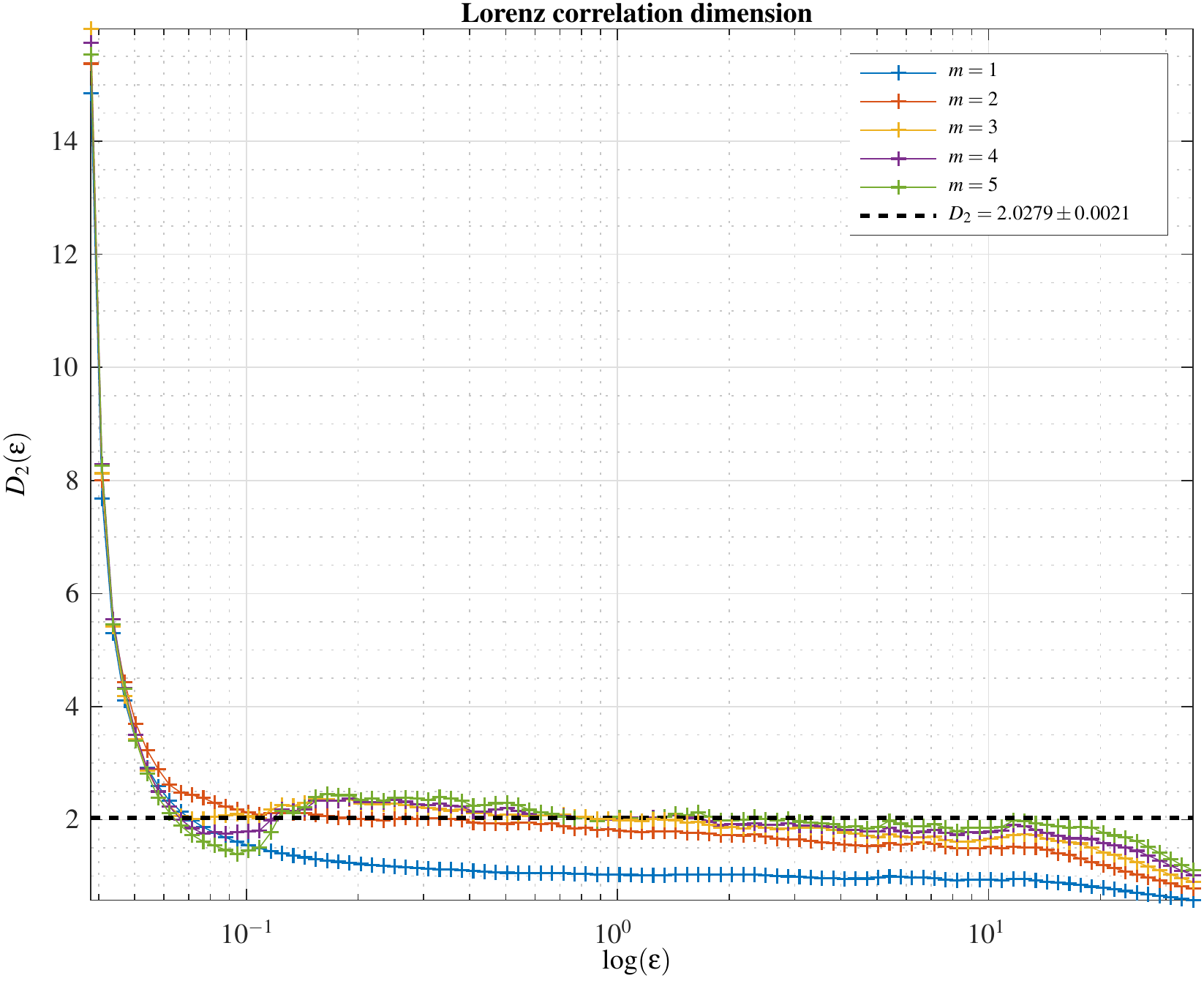}
\label{fig:F08b}}
\subfloat[]{\includegraphics[scale=0.32]{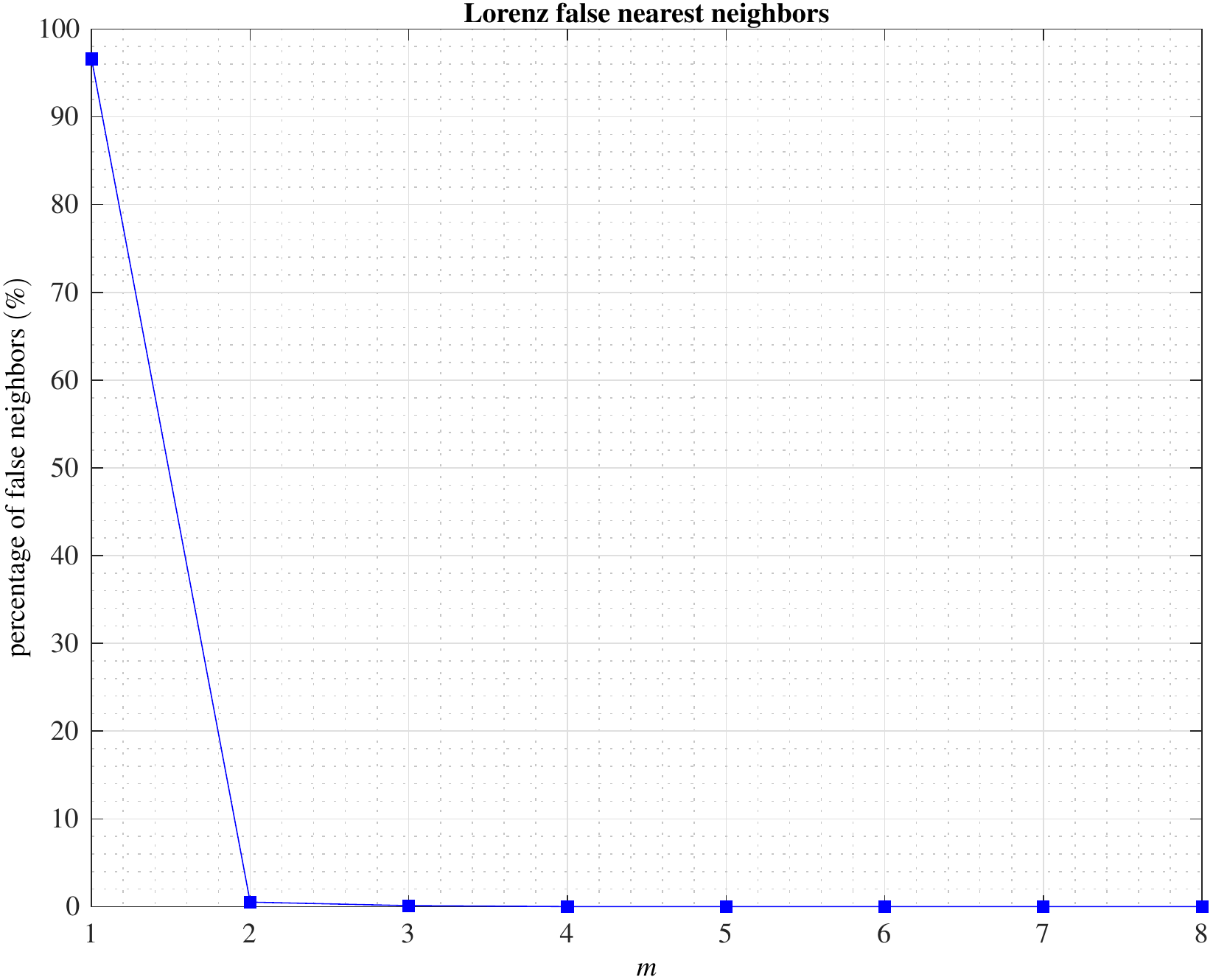}
\label{fig:F08c}}
\caption{From the dynamic variable $x$ of the Lorenz attractor: (a) Principal Component Analysis (PCA). The inner box clearly shows how the dynamic behavior of the attractor can be characterized with only three directions; (b) Correlation dimension ($D_{2}$). It can be seen how from $m=3$ the correlation dimension is confined to a constant value in an intermediate and reduced range of scales $\varepsilon$; (c) False nearest neighbors. From $m=3$ the number of false nearest neighbors is almost zero.}
\label{fig:F08}
\end{figure*}

\subsubsection{Principal Component Analysis (PCA)}

A first approach to the estimation of the parameter $m$ comes from the Principal Component Analysis (PCA), an application of linear algebra. Fig. \ref{fig:F08a} shows the results for Lorenz attractor. This method can help to extract relevant information from seemingly complex data, a priori reducing the number of dimensions needed to characterize the dynamics present in the data, that is, to identify the basis that best represents a noisy data set, detecting those redundant dimensions that record the same dynamic information\cite{Broomhead1986}. This basis, to a certain extent, filters the noise and improves the recovery of the hidden dynamics in data. In all, this linear composition aims to find the smallest subspace (hyperplane) that contains roughly the attractor.

The PCA method takes up an initial reconstruction dimension $m$ much higher than the one that supposedly characterizes the attractor. The experimenter intuition plays an important role in deciding on a greater or lesser initial value of $m$. It points the directions of the reconstruction space that show more significant variations (variances) in the data, discarding those other directions in which small variations may come from fluctuations of the coupled noise on data. Discarding those less important directions, apart from reducing the effect of noise, particularly white noise, makes it possible for appearance of a more simplified dynamics from a very high dimensional space. Hence, it enables a dimensional reduction. The main disadvantage of this method is subjective nature when determining the value of $m$. With real data, or even with an unfortunate initial choice of $\tau$ and $m$, the differences between the variances of the different dimensions are not always so evident, and it is necessary to define a somewhat arbitrary threshold that classifies the dimensions as primary or secondary. Furthermore, there is no guarantee that the reconstruction will always be optimal, since the reconstruction method relies on a non-parametric analysis, and sometimes it is unable to distinguish between a chaotic signal and the noise itself when they have a similar power spectrum\cite{Fraser1989}.

\subsubsection{Correlation dimension}

The correlation dimension involves the most usual measure of dimension, mainly because of its computational efficiency. It is based on spatial correlation\cite{Grassberger1983}. The general procedure includes point counting within a distance $\varepsilon$, evaluated for each of the points that make up the state space. The normalized total number of points, for a particular distance $\varepsilon$, is called the correlation sum, the estimator of the correlation integral,

\begin{equation}\label{eqn:E08}
C_{\varepsilon}=\frac{\text{points within a distance $\varepsilon$}}{N(N-1)},
\end{equation}
where $N$ is the time series length. For large enough $N$,

\begin{equation}\label{eqn:E09}
C_{\varepsilon}=\underset{N\rightarrow \infty}{\lim}\frac{\text{points within a distance $\varepsilon$}}{N^2},
\end{equation}
which applies for a $N$ value of several hundred data; according to the mathematical formalism, 

\begin{equation}\label{eqn:E10}
C_{\varepsilon}=\underset{N\rightarrow \infty}{\lim}\frac{1}{N^2}\sum_{i=1}^{N}\sum_{j=1}^{N}\Theta\left(\varepsilon-\left\Vert\text{\textbf{s}}_{i}-\text{\textbf{s}}_{j}\right\Vert\right),
\end{equation}
where $\Theta$ is the Heaviside step function, so, $\Theta(x)=1$, for $x\geq 0$, and $\Theta(x)=0$, for $x< 0$, and $\left\Vert\cdot\right\Vert$ symbolizes the maximum norm of one vector. 

The equation \eqref{eqn:E10} is computed for different values of $\varepsilon$, with the vectors obtained from the equation \eqref{eqn:E04} for each particular case of $m$. If the data obtained, $C_{\varepsilon}$ versus $\varepsilon$, are arranged in a straight line in a log--log plot, especially for the intermediate values of $\varepsilon$, that means that the correlation sum follows a power law. For extreme values of $\varepsilon$, the statistical estimation of $C_{\varepsilon}$ is not reliable, and, therefore, the points obtained usually move away from the straight line. Consequently, a straight line, in a log--log plot and for a central region of values of $\varepsilon$, suggests a power law, which, in this case, obeys

\begin{equation}\label{eqn:E11}
C_{\varepsilon}\propto \varepsilon^{\text{dim}_{\text{C}_{\varepsilon}}},
\end{equation}
where the exponent $\text{dim}_{\text{C}_{\varepsilon}}\equiv D_{2}(\varepsilon)$, often a non-integer number, denotes the slope of the line, and it is called correlation dimension. 

For different values of $m$ are generally obtained different values of the slope of the line. The minimum correlation dimension value, for which additional increments of $m$ do not clearly modify its value, implicitly defines the appropriate reconstruction dimension, say, $m>2\left\lceil D_{2}\right\rceil$, in accordance with Takens's theorem\cite{Takens1981}, so that the attractor can fully deploy its dynamics (see Fig. \ref{fig:F08b}). If the correlation dimension grows continuously with each value of $m$ initially chosen, the data suggest a random behavior.

\subsubsection{False nearest neighbors}

In a reconstruction space of very low dimension, two points appear to be closer to each other than they are. Two points are considered as real neighbors if the distance between them remains constant as the reconstruction dimension increases. Conversely, the distance between false neighbors continues to increase as long as the reconstruction dimension remains too low.

The underlying principle behind the method is to look for points in the time series that are neighbors in the reconstruction space, but that should not be, as their future time evolution is very different\cite{Liebert1991, Kennel2002}. The distances between all points, for different consecutive reconstruction space dimensions, $m$-dimensional and $(m+1)$-dimensional spaces, must be estimated. If the ratio between both distances is higher than a threshold $r$, it says that the neighbors are false neighbors.

If the standard deviation of the data is $\sigma$, and it uses the maximum norm, for computation speed reasons, the percentage of false neighbors $\chi$ amounts to

\begin{widetext}
\begin{equation}\label{eqn:E12}
\chi(r)=\frac{\sum_{n=1}^{N-m-1}\Theta\left(\frac{\left\Vert\mathbf{s}_{n}^{(m+1)}-\mathbf{s}_{k(n)}^{(m+1)}\right\Vert}{\left\Vert\mathbf{s}_{n}^{(m)}-\mathbf{s}_{k(n)}^{(m)}\right\Vert}-r\right)\Theta\left(\frac{\sigma}{r}-\left\Vert\mathbf{s}_{n}^{(m)}-\mathbf{s}_{k(n)}^{(m)}\right\Vert\right)}{\sum_{n=1}^{N-m-1}\Theta\left(\frac{\sigma}{r}-\left\Vert\mathbf{s}_{n}^{(m)}-\mathbf{s}_{k(n)}^{(m)}\right\Vert\right)},
\end{equation}
where $\mathbf{s}_{k(n)}^{(m)}$ is the nearest neighbor to $\mathbf{s}_{n}$ in the $m$-dimensional reconstruction space. The subscript $k(n)$ indexes the time series element, with $k(n)\neq n$, for which $\left\Vert\mathbf{s}_{n}-\mathbf{s}_{k(n)}\right\Vert$ is minimal.
\end{widetext}

The first term of the numerator in equation \eqref{eqn:E12}, within the summation, is equal to one, if the nearest neighbor is false, that is, if the distance increases by a factor greater than $r$ when the reconstruction dimension is increased by one, from $m$ to $m+1$. The second term drops those pairs of points whose initial distance is already greater than $\sigma/r$, since, by definition, they can not be false neighbors, as, on average, the points cannot be further away than $\sigma$. Therefore, in the calculation, these points should not be taken and, thus, appear in the denominator's normalization factor.

The right reconstruction dimension is that for which the percentage of false neighbors falls to approximately zero, as shown in Fig. \ref{fig:F08c}. Once it reaches the dimension, it assumed that the attractor embraces its true spatial configuration, at least from a topological perspective. The results may depend on the chosen delay $\tau$. In any case, if it seeks to be able to differentiate the chaos from the noise, it is essential to validate the results using surrogate data testing\cite{Theiler1992}.

\begin{figure*}[ht!]
\centering
\subfloat[]{\includegraphics[scale=0.24]{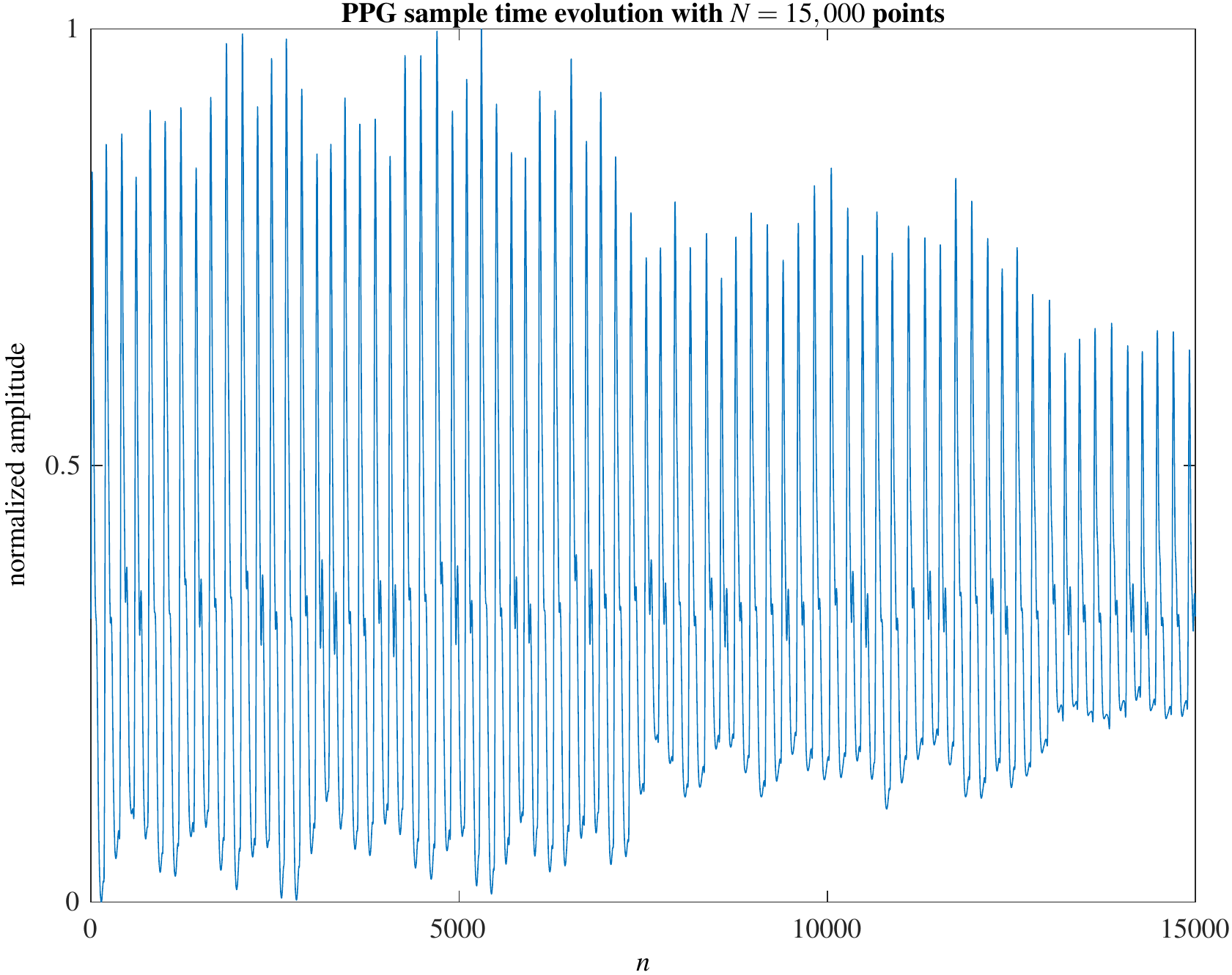}
\label{fig:F09a}}
\subfloat[]{\includegraphics[scale=0.24]{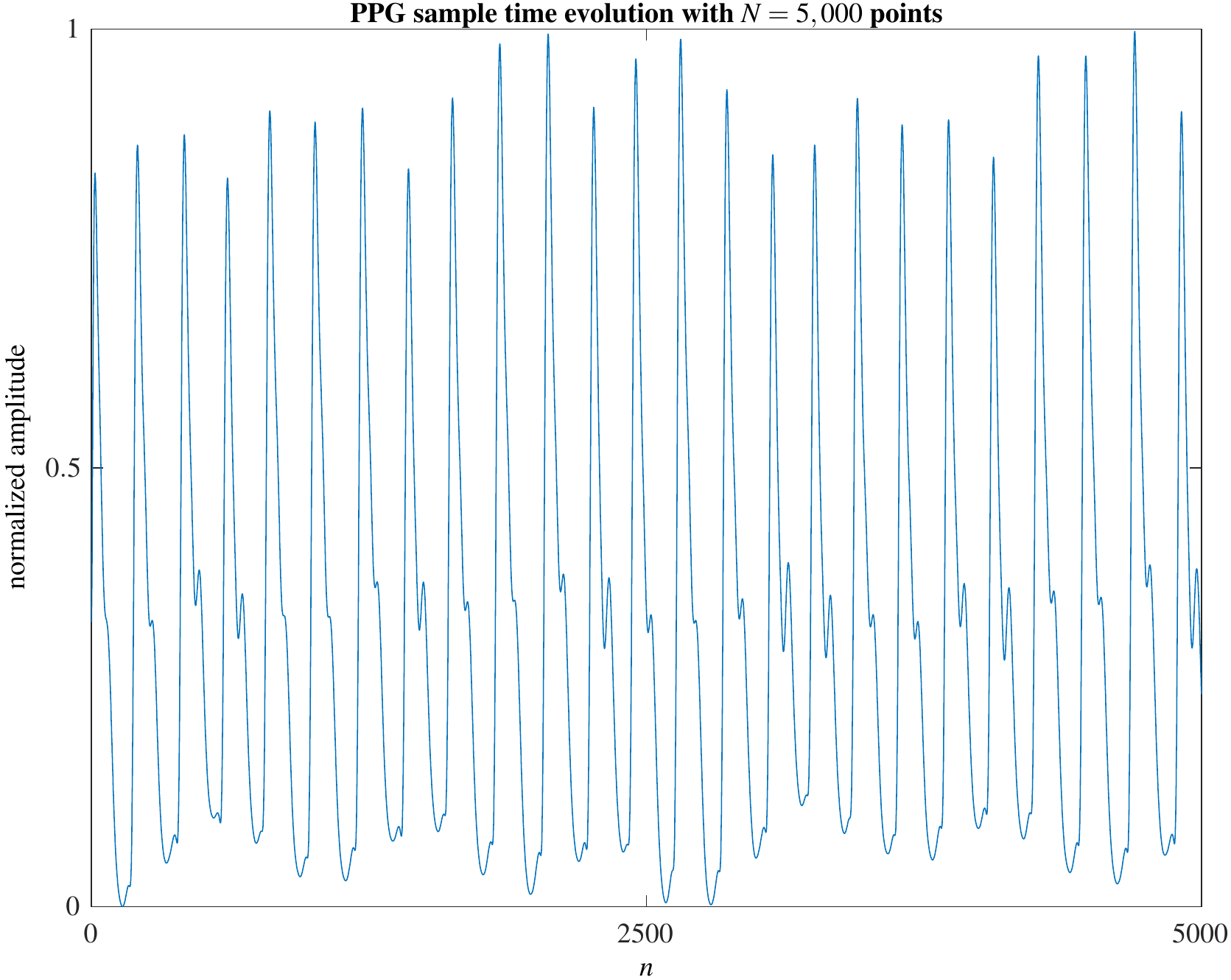}
\label{fig:F09b}}\hfil
\subfloat[]{\includegraphics[scale=0.24]{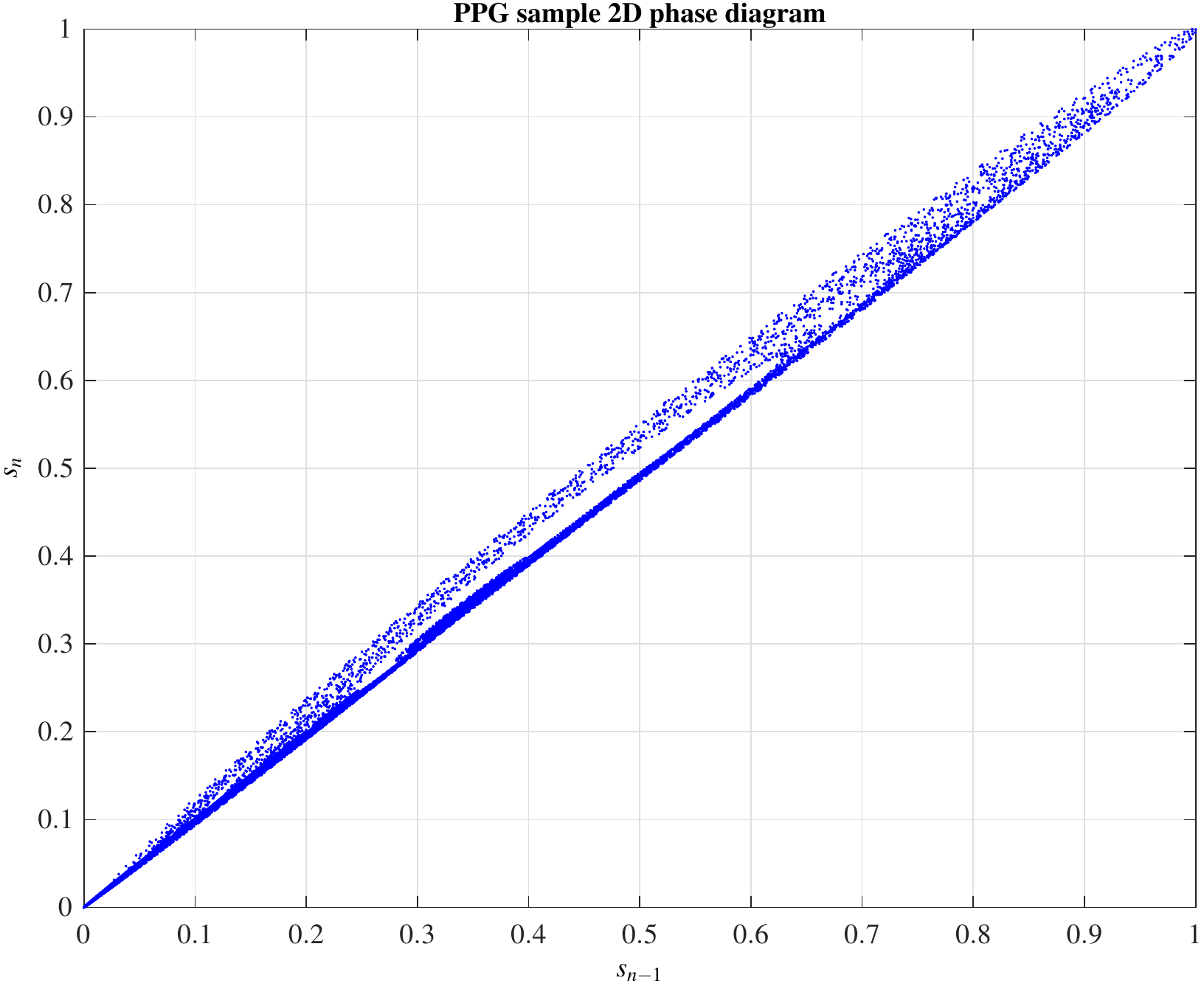}
\label{fig:F09c}}
\subfloat[]{\includegraphics[scale=0.24]{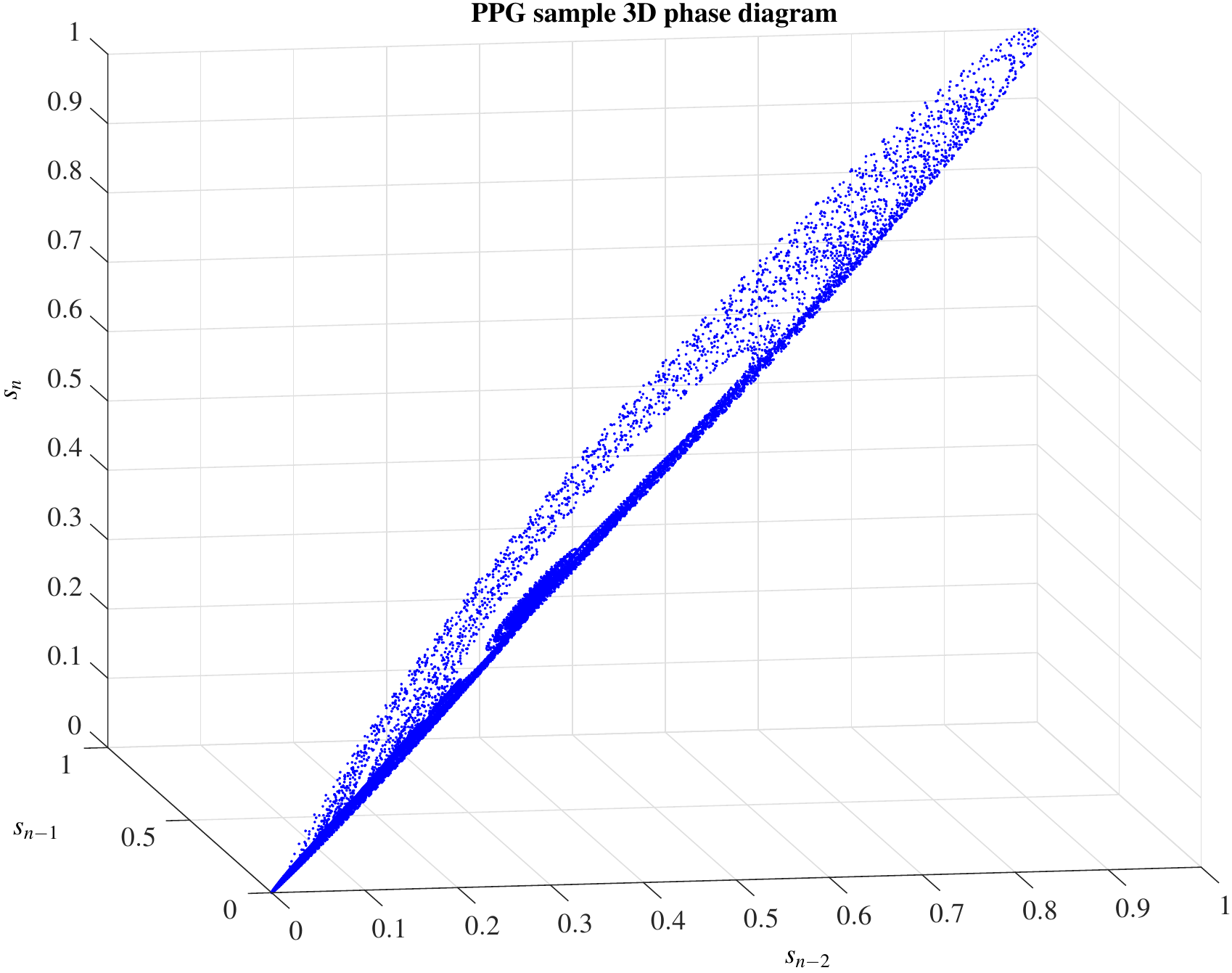}
\label{fig:F09d}}
\subfloat[]{\includegraphics[scale=0.24]{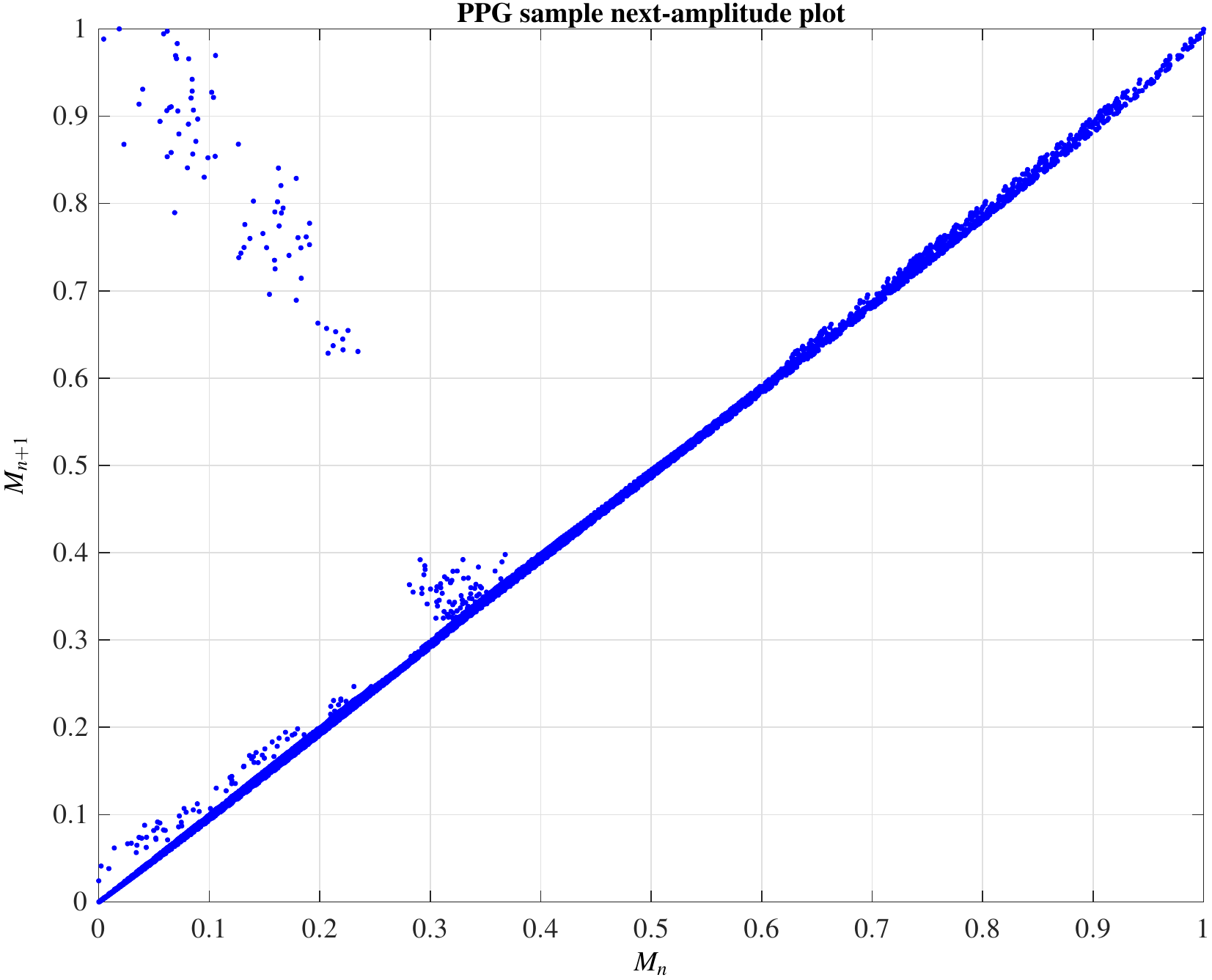}
\label{fig:F09e}}
\subfloat[]{\includegraphics[scale=0.24]{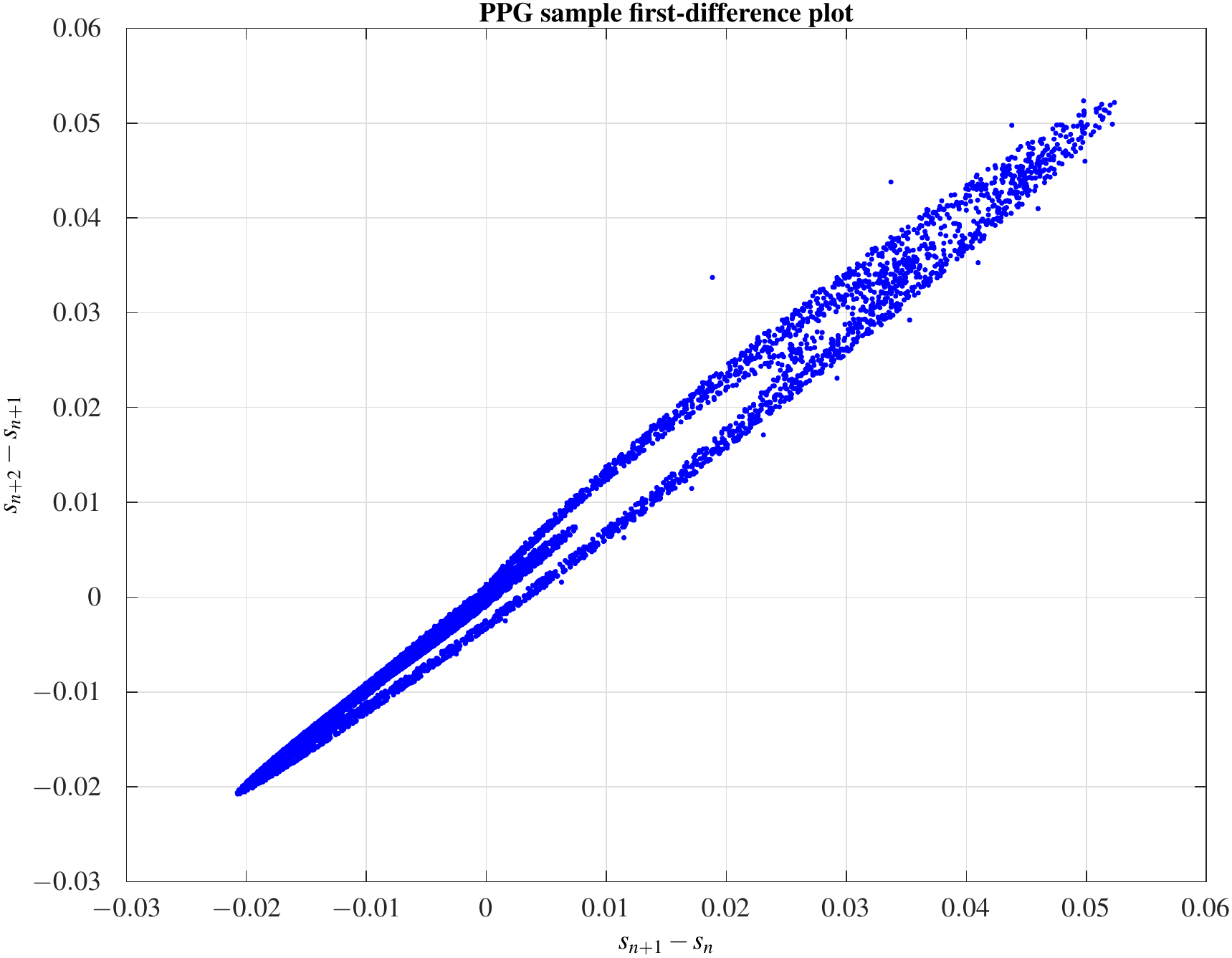}
\label{fig:F09f}}\hfil
\subfloat[]{\includegraphics[scale=0.24]{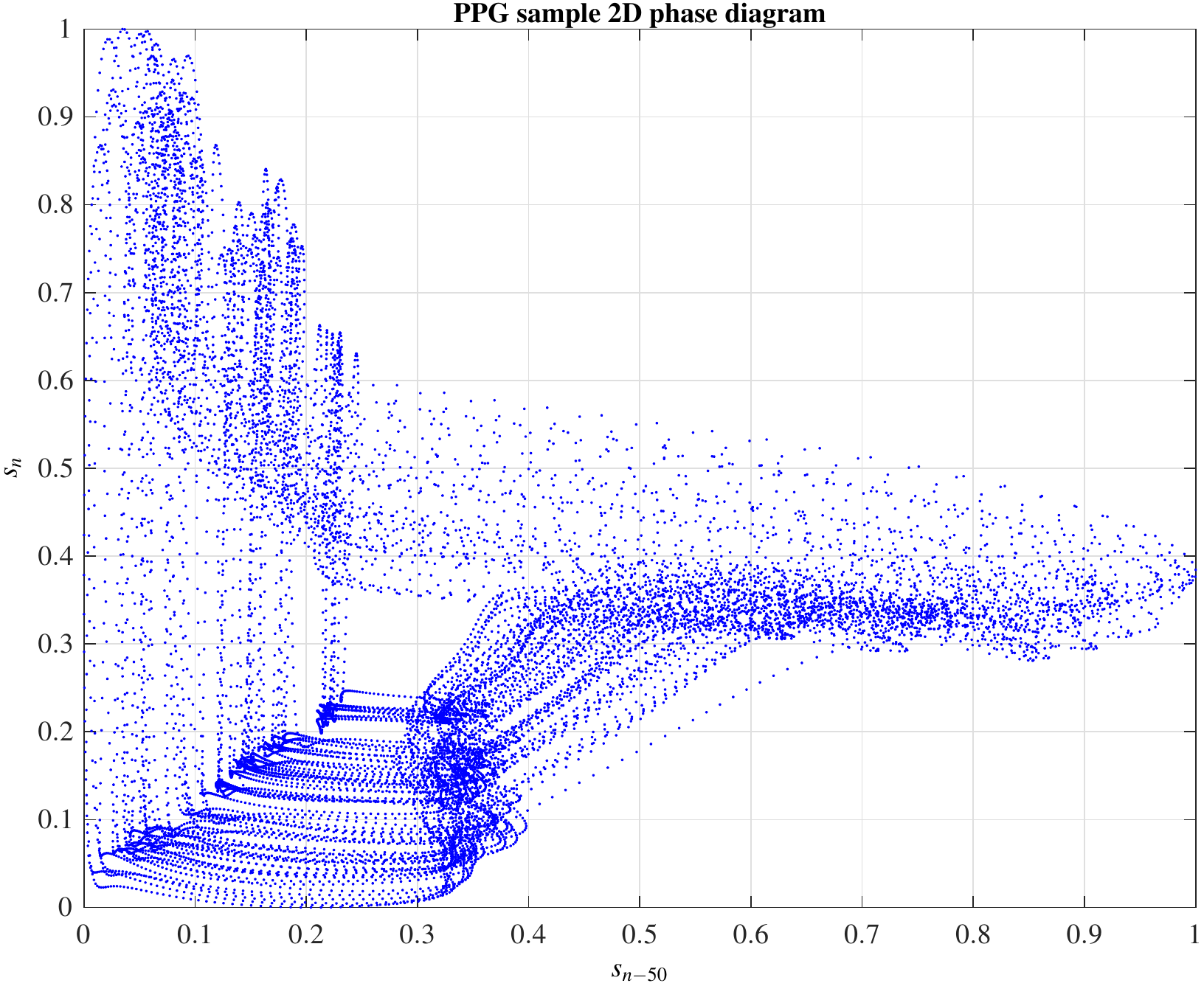}
\label{fig:F09g}}
\subfloat[]{\includegraphics[scale=0.24]{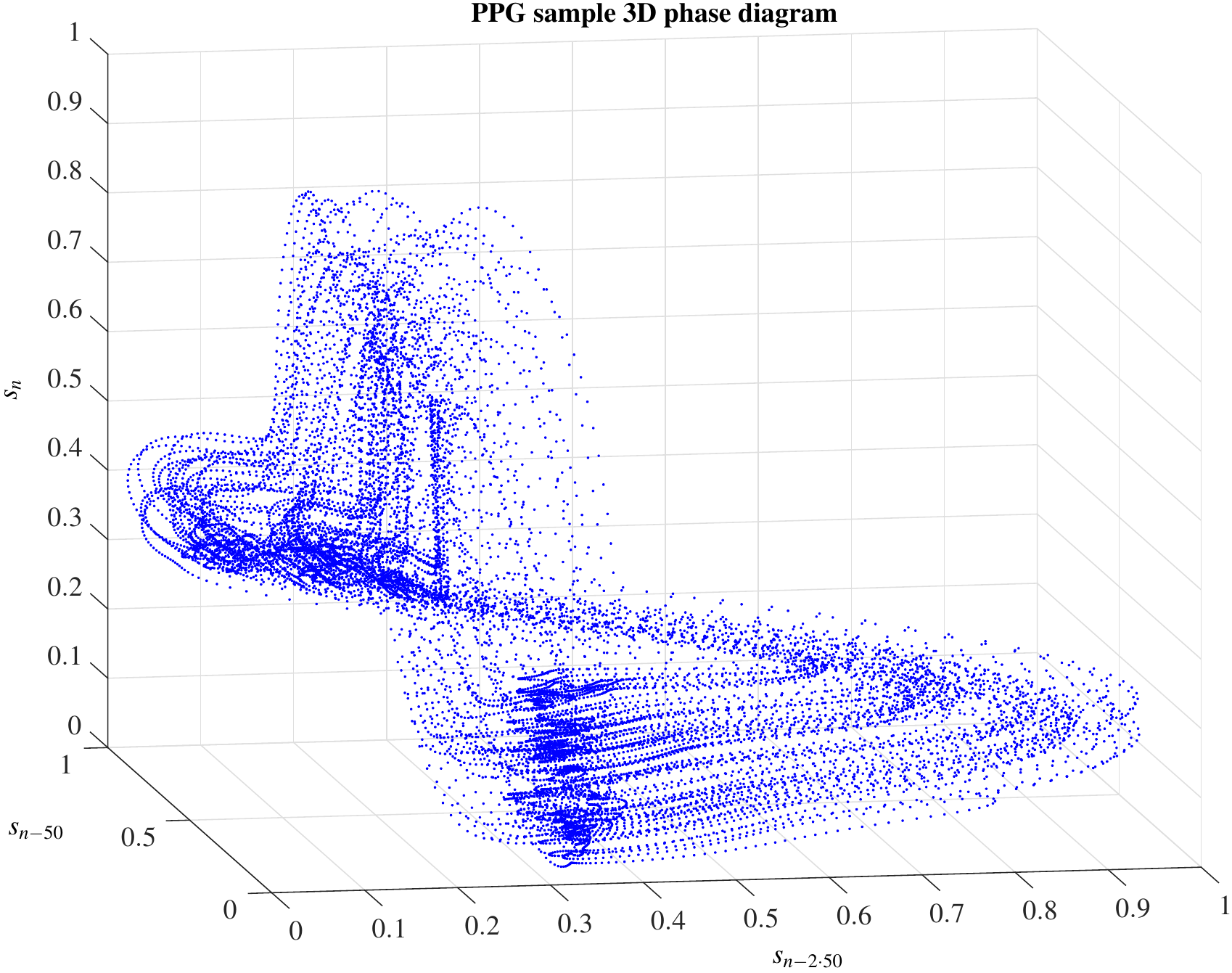}
\label{fig:F09h}}
\subfloat[]{\includegraphics[scale=0.24]{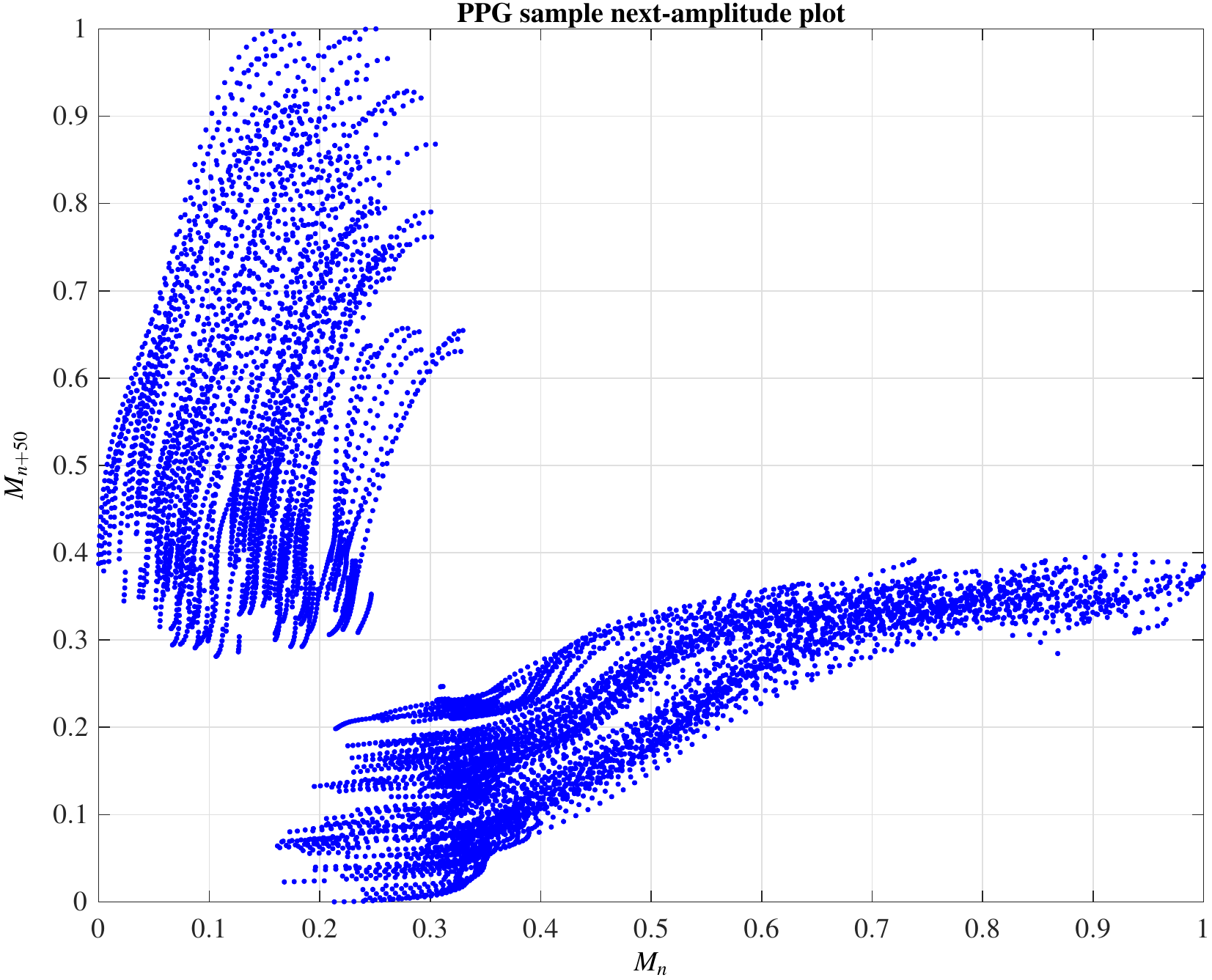}
\label{fig:F09i}}
\subfloat[]{\includegraphics[scale=0.24]{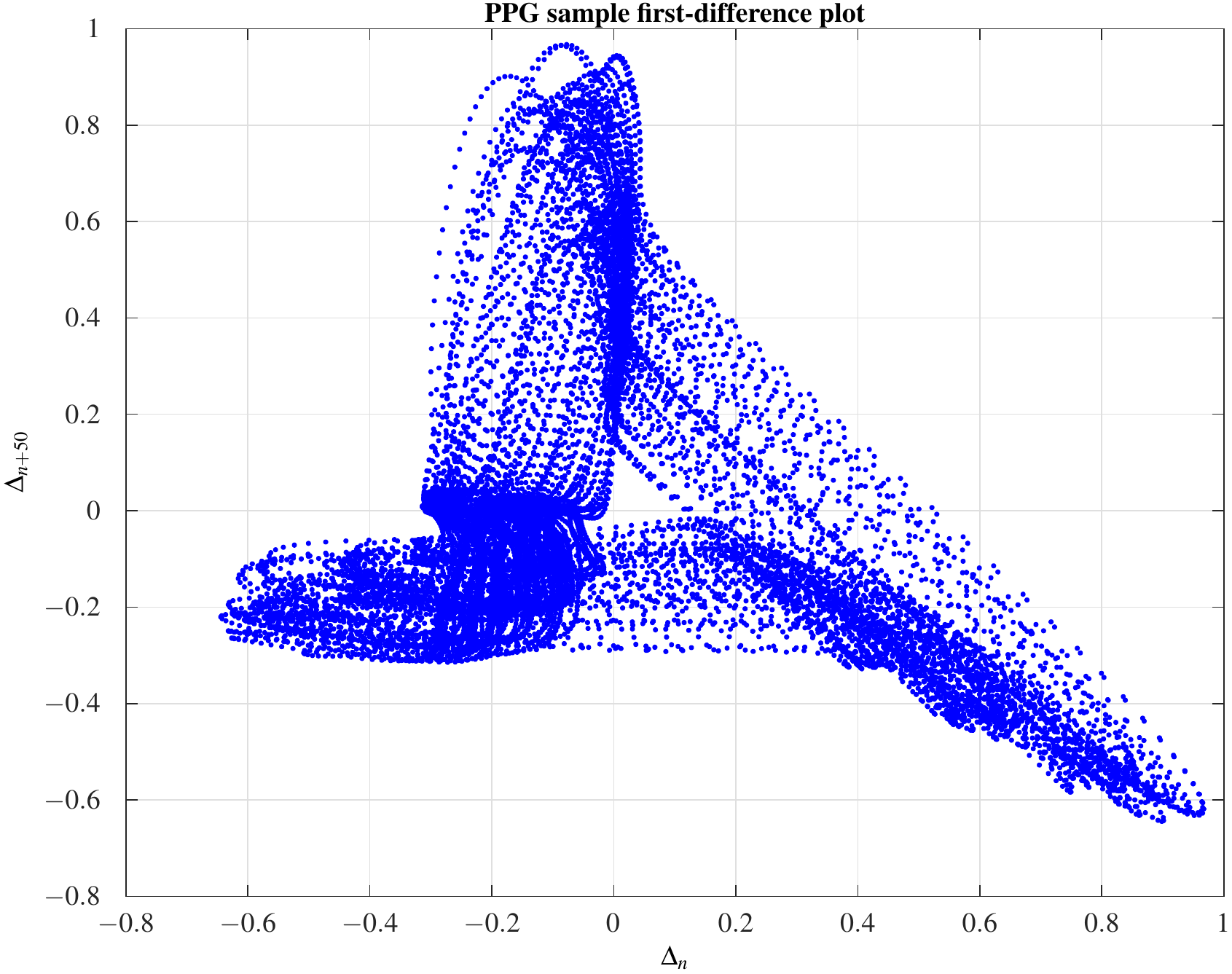}
\label{fig:F09j}}
\caption{From a sample PPG signal: (a) 1 minute PPG recorded signal of a healthy young subject; (b) the first 20 seconds of (a); (c)-(f) 2D and 3D phase diagrams, successive-maxima and difference plots for $\tau=1$; (g)-(j) as in the previous cases but now for $\tau=50$.}
\label{fig:F09}
\end{figure*}

\section{Biological data}\label{sec:BIO}

The health monitoring by non-invasive means has attracted the attention of medical specialists for some years now since it allows to advance preliminary diagnoses on possible pathological dysfunctions, whether chronic or transient. Various scientific evidences sustain the relationship between the biological signals generated by the human body and the health status of the individual, as skin temperature (ST), electrodermal activity (EDA), pulse wave or photoplethysmography (PPG), electrocardiography (ECG), electromyography (EMG), respiration (Resp.), pupil diameter (PD), electroencephalography (EEG), blood pressure, among others. The information provided by all these biological signals depends on the knowledge available about the underlying physiological processes.

The study of the dynamics of all these biological signals will help to  understand better the physiological system that generates them. Also, it would help to get an insight about how different dynamic variables couple, as, for instance, the heart and respiratory frequencies, in order to keep the physiological system in a perfect condition, preventing its deterioration towards a pathological state. In this paper, we focus only on a single biological signal, the photoplethysmographic (PPG) signal; forward publications will describe the results for more biological signals.

\subsection{Results on the PPG signal case study}

We have chosen the PPG signal because it is easily accessible. A pulse oximeter consists of a light emitter and a photodetector that collects and records (pulse or PPG signal) the loss---scattering and absorption---that a beam of light undergoes when it passes through or is reflected, a human tissue\cite{Cano-Garcia2018}. It allows detecting blood volume changes in the microvascular bed of tissue---in our case, the middle finger of the left hand---, obtaining valuable information about the cardiovascular system and, on the whole, about the cardiorespiratory system. Given the simplicity of its non-invasive accommodation, in addition to its low cost, a pulse oximeter is very useful in biomedical applications for clinical\cite{Allen2007} and sports environments\cite{Alves2016}. As with other biological signals, characteristics extracted of the PPG signal allow that to no small extent identify ideal health conditions and their possible deviations. Thus, indicators associated with different pathologies could be established, which anticipate its severity according to the causes that gave rise them. Typically, these indicators in the case of PPG signal were based on the morphology of the signal rather than on its dynamics\cite{Elgendi2018, Allen2007}; we think that by studying dynamic aspects of the PPG signal, the physiological system that generates it can be better understood.

With the aim of examining the PPG signal dynamics, we use the PPG signals (to show in this paper only five individuals chosen randomly) from a total of 40 students, between 18 and 30 years old and a non-regular consumers of psychotropic substances, alcohol or tobacco, selected to participate in a national research study\cite{Aguilo2015,Arza2018}; the five PPG signals show in this paper are the same as those used in previous research that confirms its predominantly quasi-periodic behavior for small timescales in healthy young people with the 0--1 test\cite{Pedro-Carracedo2019}. To show more than five subjects will not clarify the proposed method. The results are similar. Remember that the fundamental frequency of the PPG signal is typically around 1 Hz, depending on heart rate (0.5--4 Hz, first and second harmonic), and respiratory activity at 0.2--0.35 Hz. We apply a Butterworth bandpass filter with cutoff frequencies at 0.1 and 8 Hz, in order to avoid high-frequency noise and to some extent, motion artifacts\cite{RAM2012}. All signals were captured from the middle finger of the left hand and sampled at a frequency of 250 Hz\cite{Aguilo2015}, say, sampling time $\Delta t=4$ ms.

\begin{figure}[ht!]
\centering
\subfloat[]{\includegraphics[width=0.47\columnwidth]{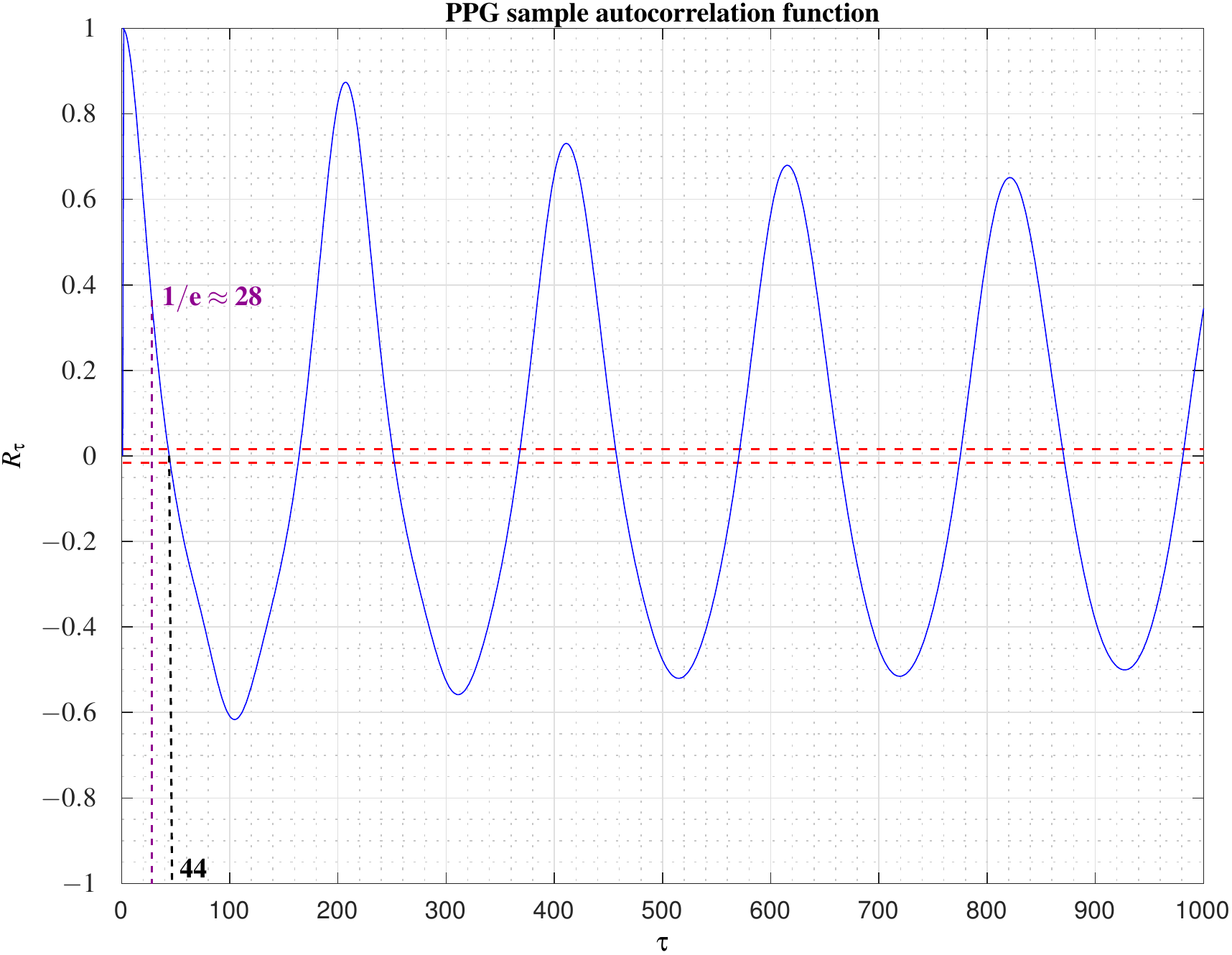}
\label{fig:F10a}}
\subfloat[]{\includegraphics[width=0.47\columnwidth]{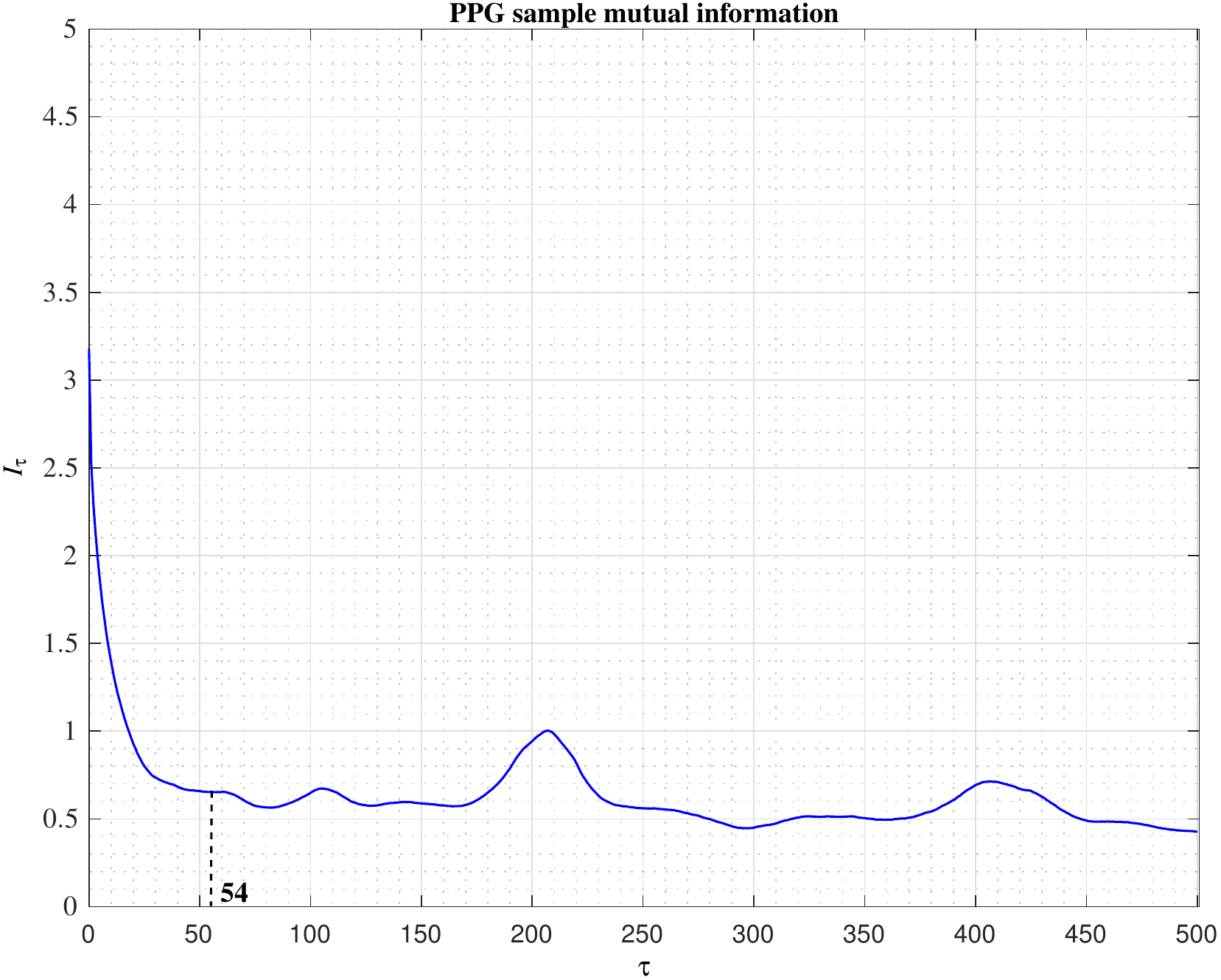}
\label{fig:F10b}}
\caption{From a sample PPG signal: (a) Autocorrelation Function (AF), where the first zero crossing is when $\tau=44$; (b) Mutual Information (MI), where the first minimum is with $\tau=54$.}
\label{fig:F10}
\end{figure}

\begin{figure*}[ht!]
\centering
\subfloat[]{\includegraphics[scale=0.32]{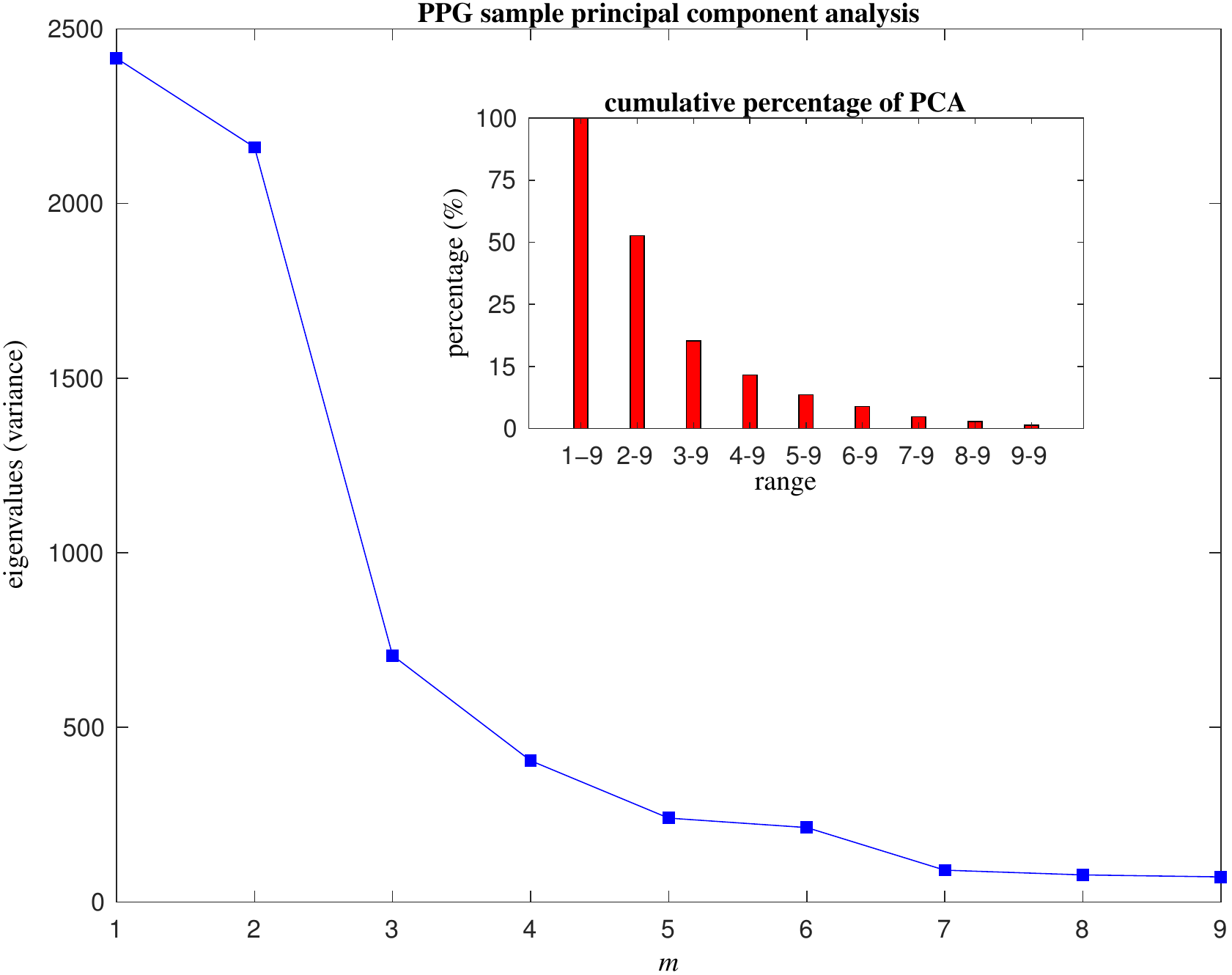}
\label{fig:F11a}}
\subfloat[]{\includegraphics[scale=0.32]{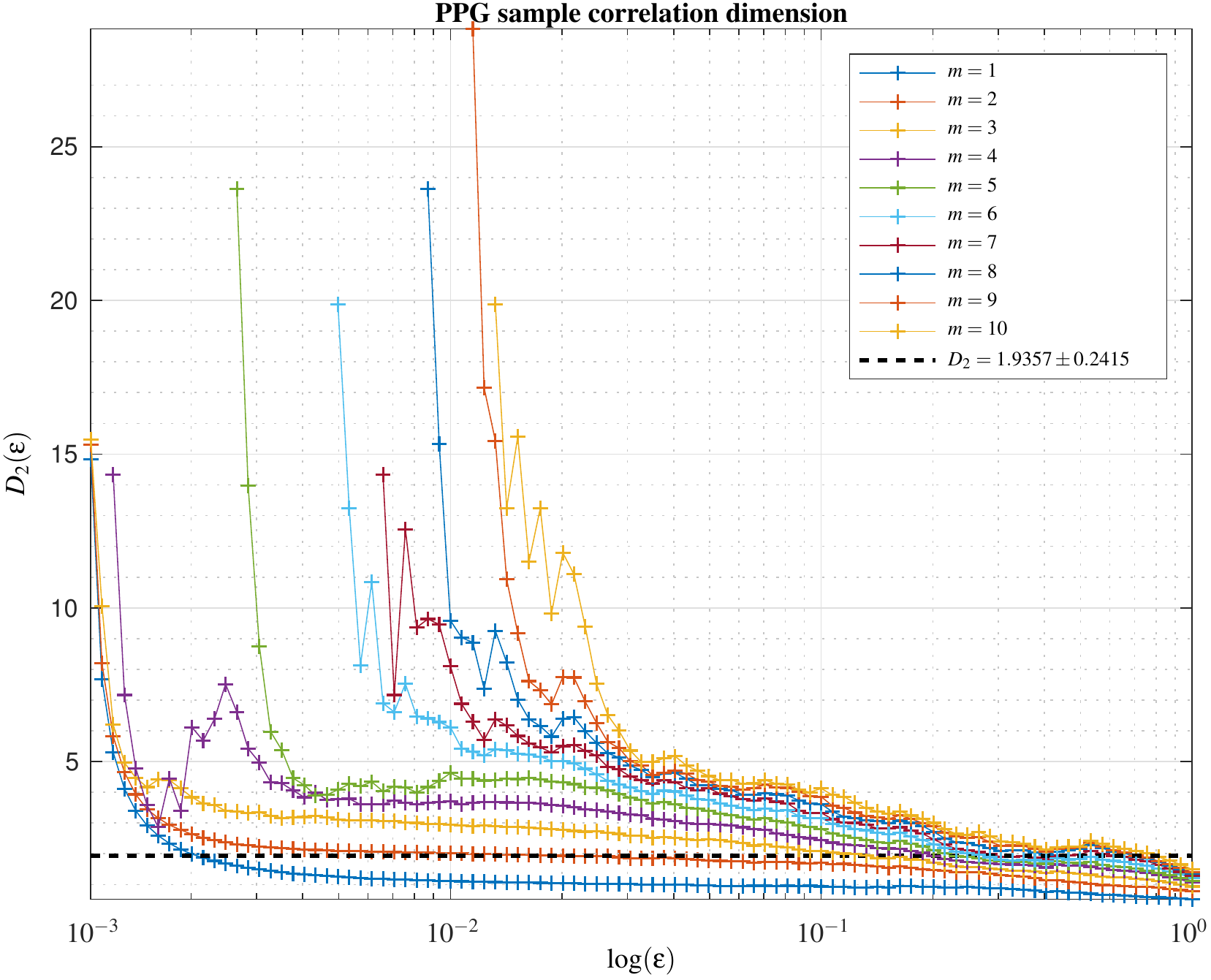}
\label{fig:F11b}}
\subfloat[]{\includegraphics[scale=0.32]{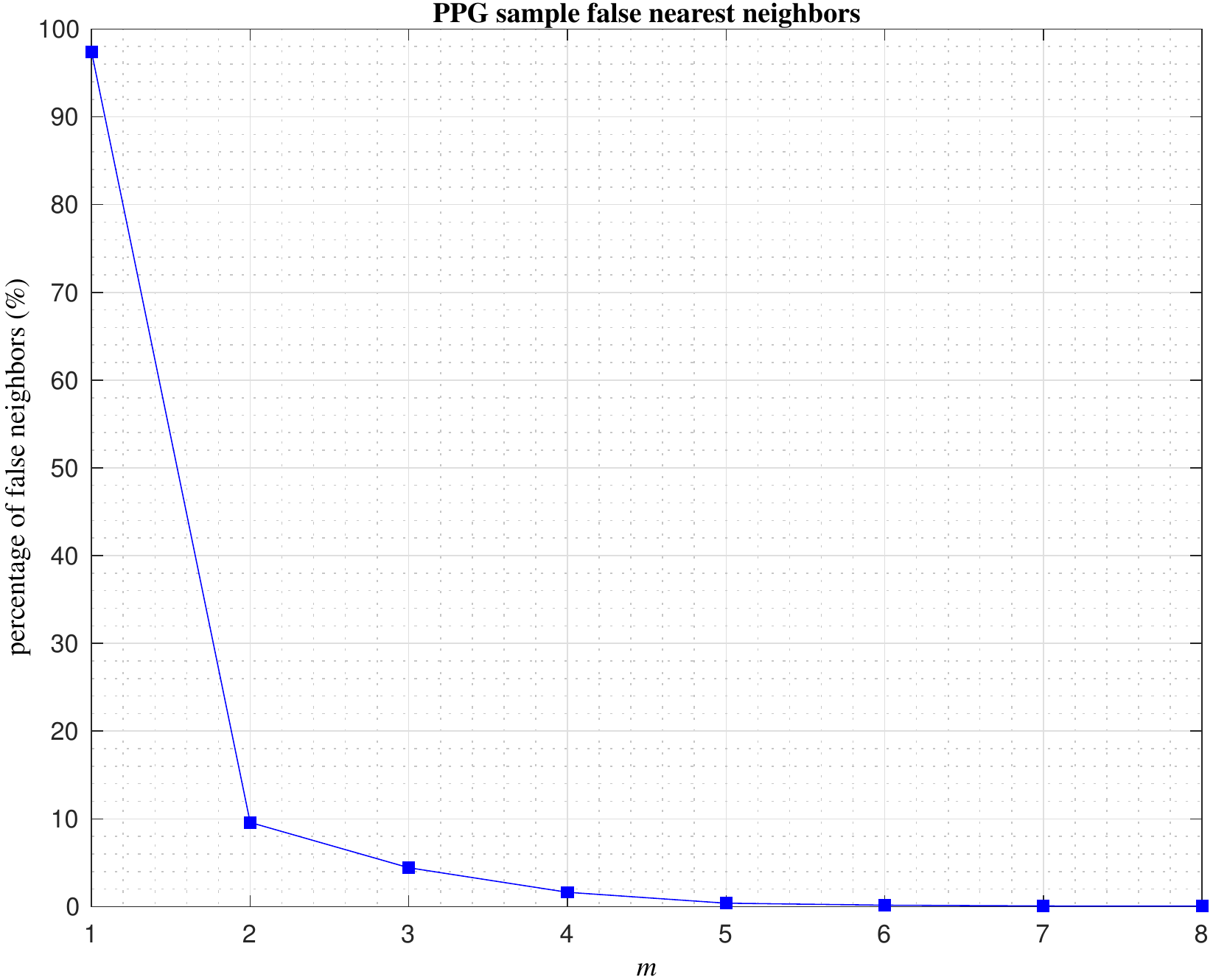}
\label{fig:F11c}}
\caption{From a sample PPG signal: (a) Principal Component Analysis (PCA). The inner box clearly shows how the dynamic behavior of the attractor can be characterized with roughly seven directions; (b) Correlation dimension ($D_{2}$). It can be seen how from $m=4$ the correlation dimension is confined to a constant value in an intermediate and reduced range of scales $\varepsilon$; (c) False nearest neighbors. From $m=5$ the number of false nearest neighbors is almost zero.}
\label{fig:F11}
\end{figure*}

The study methodology of the PPG signal to finally get the phase space reconstruction and draw interesting conclusions about the underlying physiological dynamics follow the next steps: first, we take 15,000 points which correspond to one minute recorded signal and study its graphical representation, considering different options for different graphic representations, as shown in Fig. \ref{fig:F09} (in the accompanying video clips we can watch the evolution of the attractor geometry for different lags, for both 2D and 3D phase diagrams), to sustain the presence of a latent dynamic structure. Second, we try to figure out the attractor more akin to the original, based on the optimal lag $\tau$, see subsection \ref{ssec:LAG}, according to the Autocorrelation Function (AF) (Fig. \ref{fig:F10a}) and the Mutual Information (MI) (Fig. \ref{fig:F10b}), and on the optimal embedding dimension $m$, see subsection \ref{ssec:EMB}, in accordance with the Principal Component Analysis (PCA) (Fig. \ref{fig:F11a}), the Correlation Dimension (D2) (Fig. \ref{fig:F11b}), and the False Nearest Neighbors method (FNN) (Fig. \ref{fig:F11c}). We apply the methodology to study 10 minutes of the PPG signal from each subject, and the results are similar to the ones shown in this paper.

We have seen in the Lorenz flow, according to the AF, that the criterion of $1/e$ is approximately similar to the MI result because the attractor is chaotic and both linear and nonlinear correlations decay rapidly, as illustrated in Fig. \ref{fig:F07}. In the PPG signal, the criterion of the first zero crossing is closer to the MI result, see Fig. \ref{fig:F10}, because the PPG signal is predominantly quasi-periodic on small timescales\cite{Pedro-Carracedo2019}, and the correlations are stronger. Either way, the criterion of the first minimum of MI is adopted because it ensures statistical independence between the values, in linear and nonlinear terms. The parameters $\tau$ and $m$ calculate for five subjects, with results from each of the methods previously described, are in Tab. \ref{tab:TP01}. The optimal $\tau$ depends on the subject considered and on the time interval in which it estimates. Probably the range of optimal $\tau$ values varies with each subject and could be a hallmark of each individual. It should make a careful study with more subjects.

\begin{figure*}[ht!]
\centering
\subfloat[]{\includegraphics[scale=0.55]{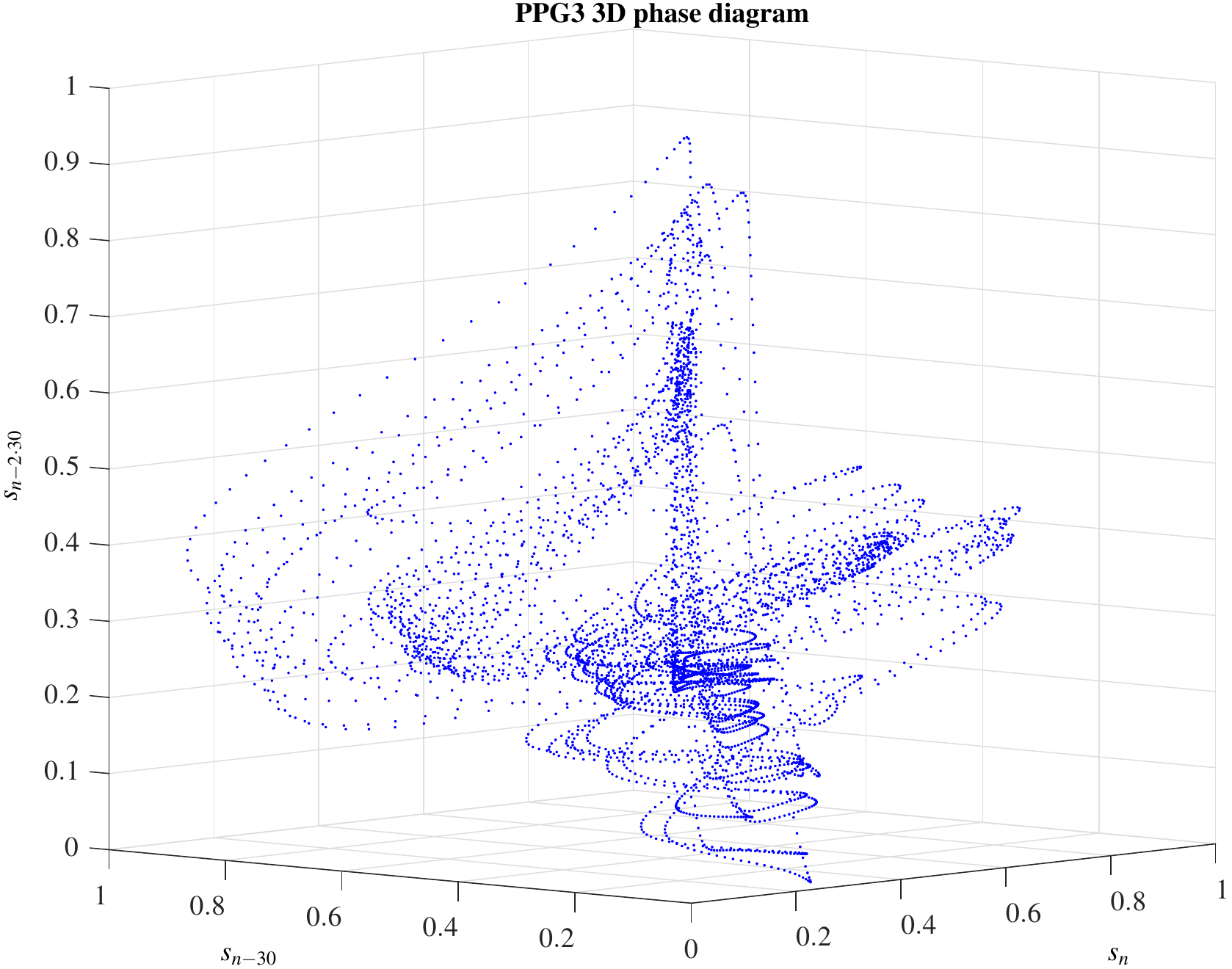}
\label{fig:F12l}}\hfil


\subfloat[]{\includegraphics[scale=0.55]{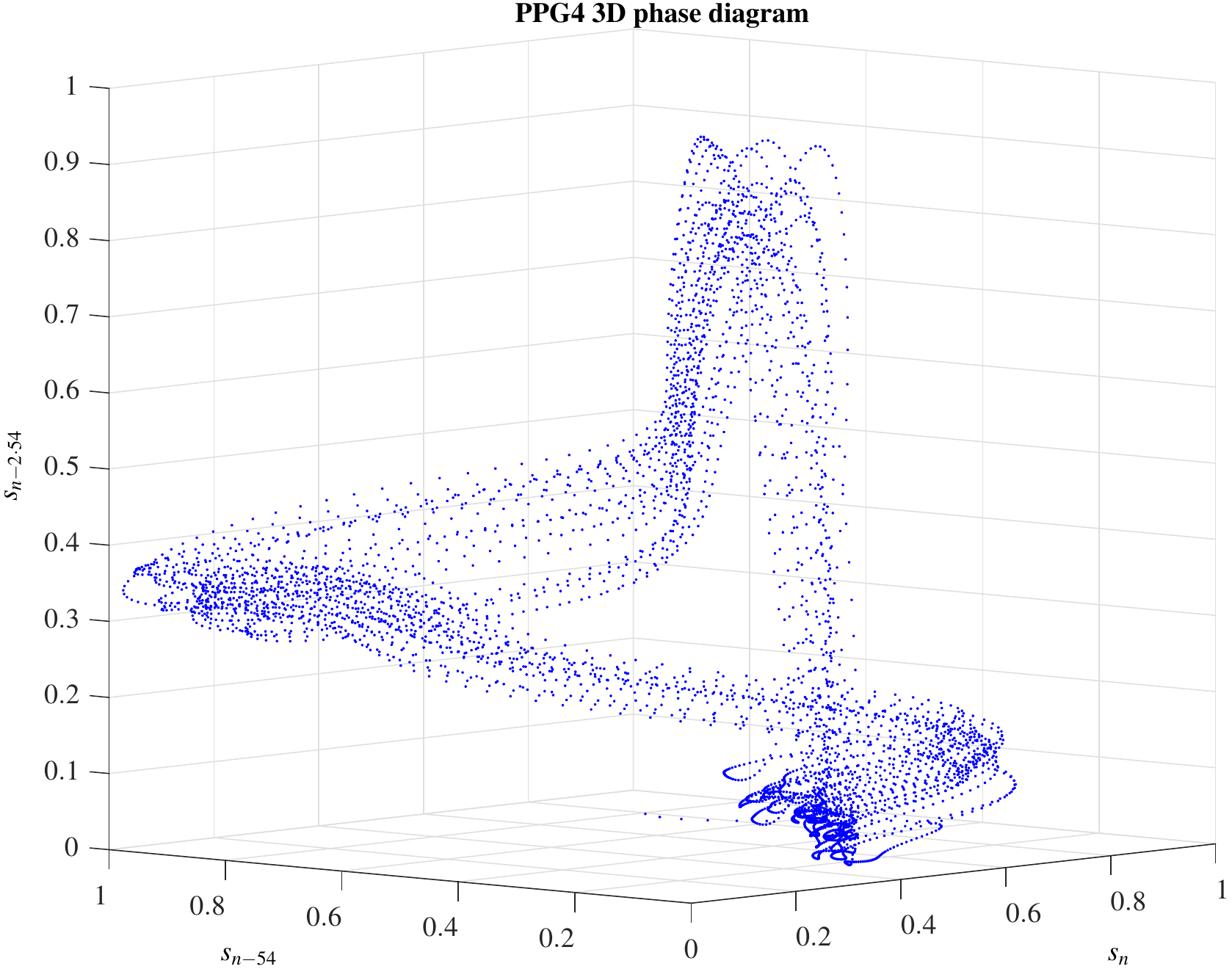}
\label{fig:F12p}}\hfil

\subfloat[]{\includegraphics[scale=0.55]{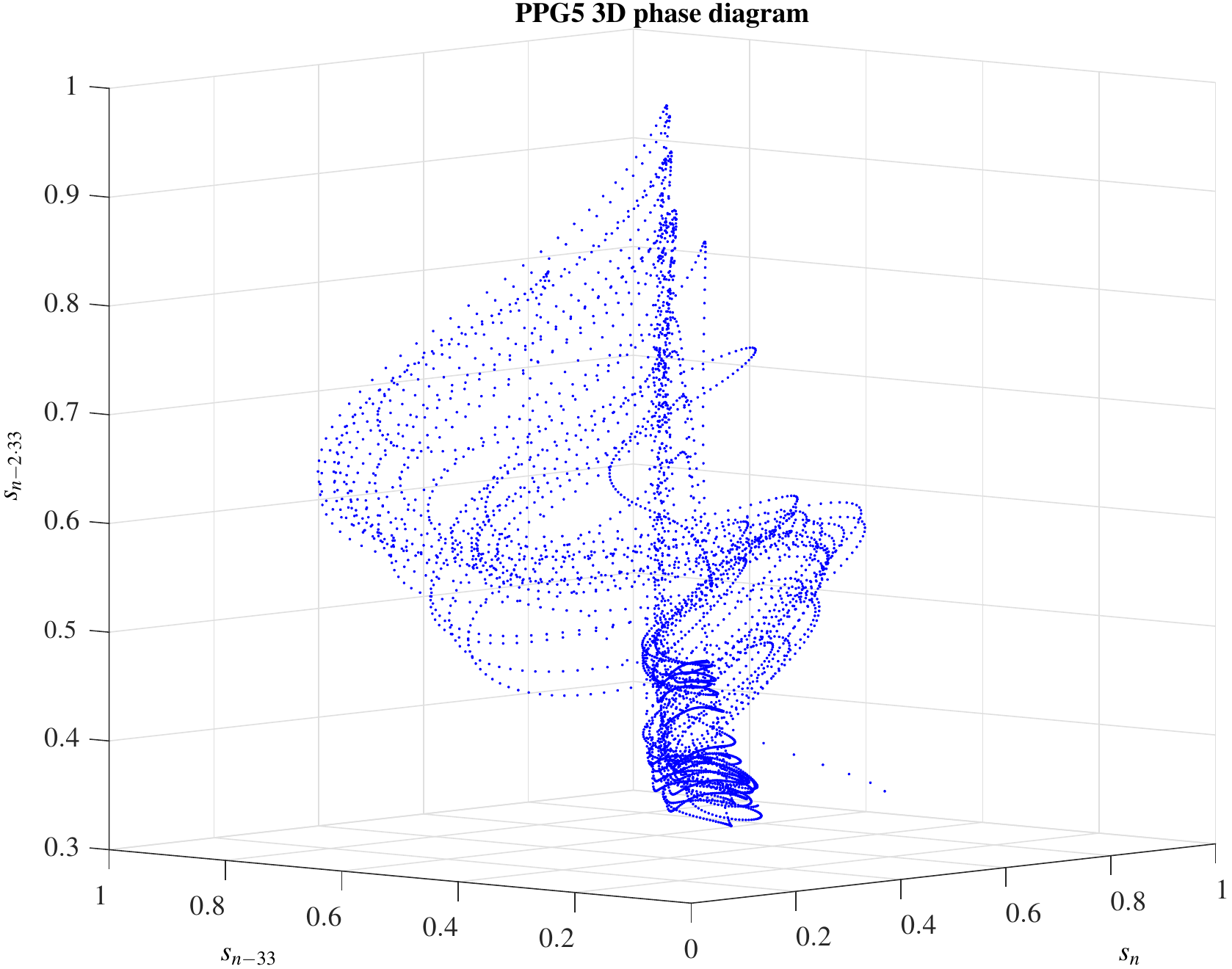}
\label{fig:F12t}}
\caption{From 3 PPG signals belonging to different subjects: (a) PPG3 3D phase diagram with $\tau=30$; (b) PPG4 3D phase diagram with $\tau=54$; (c) PPG5 3D phase diagram with $\tau=33$. For reasons of constraints relating to available space the phase diagrams of more individuals are not shown.}
\label{fig:F12}
\end{figure*}

Once we have the parameter $\tau$ for each subject, calculated by the MI criterion, we now proceed to the calculation of parameter $m$. Fig. \ref{fig:F11} visually represent the results for a sample PPG signal, which corresponds to subject number 4 of the Tab. \ref{tab:TP01}. We see that multivariate analysis, such as the PCA linear method, does not provide the ideal value of $m$, although it is not unreasonable, mostly because physical systems do not usually have very high dimensions. Correlation dimension gives the value of $m>2\left\lceil D_{2}\right\rceil$, that, except for the first subject, amounts to a value of $m=5$. For the FNN method, all calculations agree with $m=5$. This later method is a good estimator of $m$ because in resorting to topological aspects of the reconstructed state space, the effect of noise on algorithmic computation is minimal. We conclude that all individuals have the same embedding dimension $m=5$, although we think that it may vary depending on the subject's physical and psychological states. Consequently, a system of five first-order differential equations describes the dynamic system that generates a typical PPG signal from a healthy young individual, in which various parameters could contribute to the mechanism of autoregulation of the underlying physiological process. Fig. \ref{fig:F12} shows how the attractors of the subjects are different for the calculated values of $\tau$.\footnote{This graph doesn't show the information of the five individuals for reasons of available space.} Ongoing studies on the effect of changes in the parameters $\tau$ and $m$ on the dynamics of the underlying physiological process will shed any light in this regard.

\setcellgapes{5pt}\makegapedcells
\begin{table}[ht]
\centering
\caption{Parameters $\tau$ and $m$ of the PPG signals, acquired from five individuals chosen at random, for the different methods already explained, namely, autocorrelation function (\textbf{AF}), mutual information (\textbf{MI}), principal component analysis (\textbf{PCA}), correlation dimension ($\mathbf{D_{2}}$) and false nearest neighbors (\textbf{FNN}).}
\label{tab:TP01}
\begin{adjustbox}{max width=\textwidth}
\begin{tabular}{| l | c | c | c | c | c |}
\hline 
\multirow{2}{*}{\large{\textbf{evaluated signal}}} & \multicolumn{2}{c|}{\large $\tau$} & \multicolumn{3}{c|}{\large $m$}\\
\cline{2-6} 
 & \textbf{AF} & \textbf{MI} & \textbf{PCA} & $\mathbf{D_{2}}$ & \textbf{FNN}\\
\hline 
subject number 1 (PPG1) & 52 & 81 & 5 & 9 & 5\\
\hline
subject number 2 (PPG2) & 37 & 35 & 6 & 5 & 5\\
\hline 
subject number 3 (PPG3) & 29 & 30 & 5 & 5 & 5\\
\hline 
subject number 4 (PPG4) & 44 & 54 & 6 & 5 & 5\\
\hline 
subject number 5 (PPG5) & 44 & 33 & 6 & 5 & 5\\
\hline 
\end{tabular}
\end{adjustbox}
\end{table}

\section{Conclusion}\label{sec:CON}

The well-known phase space reconstruction serves as a starting point for the analysis or modeling of the dynamics of physiological systems reconstructed from a single biological signal. The dynamic that describes a photoplethysmographic signal is deterministic. The embedding dimension $m$, for healthy young subjects, is five, regardless of the subject and the time interval estimation. The optimal lag or delay time $\tau_{\text{opt}}$ depends on the subject and calculated time interval. A further study, with individuals of different ages and with different proven pathologies, will be carried out to confirm if the embedding dimension remains the same. In addition, it will be thoroughly examined if the variation of $\tau$, even of $m$, in each individual is an indicator of the mood and physical status in which he or she is, either as a result of a sporadic situation, such as an episode of stress, a more severe disease or the age-related evidence for biological and physiological decline. We describe, in some detail, the most usual and clear methodology to calculate the phase space reconstruction because we have found that in its application to biological signals is not well understood at the physiological level, and its discriminant potential in the clinical setting no sufficiently exploit. In this sense, its effectiveness could be corroborated with the most modern state space reconstruction techniques that are less heuristic and with a more consolidated mathematical formalism\cite{Pecora2007}.

\begin{acknowledgments}

The authors would like to thank Life Supporting Technologies Group
(LST-UPM) for taking part in project FIS-PI12/00514, from MINECO. Also, they
want to thank especially the support given by Ra\'ul Dur\'an D\'iaz, from the
University of Alcal\'a (UAH) and the Spanish National Research Council
(CSIC), providing with a cluster of servers with which to execute some of
the computational algorithms of this work. The cluster consists of ten
computing nodes, with a total of 64 cores and Intel Xeon architecture, and
166 GB of memory.

\end{acknowledgments}

\bibliography{Paper2}

\end{document}